\documentclass[11pt]{article}
\usepackage{hyperref}
\usepackage{amsmath}
\usepackage{amssymb}
\usepackage{mathtools}
\usepackage{slashed}
\usepackage[numbers,sort&compress]{natbib}

\usepackage{mathrsfs}
\usepackage{cleveref}
\usepackage[letterpaper,margin=1in,bottom=1in]{geometry}
\usepackage{float} 
\usepackage{parskip} 
\usepackage{tabulary} 
\usepackage{color} 
\usepackage{soul} 
\usepackage{subfigure}
\usepackage{graphicx}
\usepackage[section]{placeins} 
\usepackage{cleveref}

\newcommand{\GeV}{~\textrm{GeV}}

\def\.4{\vspace{-.5cm}}

\newcommand{\cO}{\mathcal{O}}
\newcommand{\cL}{\mathcal{L}}
\newcommand{\bP}{\mathbb{P}}

\def\beq{\begin{equation}}
\def\be{\begin{equation}}
\def\beqn{\begin{eqnarray}}
\def\ee{\end{equation}}
\def\eeq{\end{equation}}
\def\eeqn{\end{eqnarray}}

\def\eV{{\rm eV}}
\def\GeV{{\rm GeV}}
\def\TeV{{\rm TeV}}
\def\cV{\mathcal{V}}
\def\K{K\"ahler~}

\author{James Halverson\footnote{j.halverson@neu.edu}\, ,~\,
Cody Long\footnote{co.long@neu.edu}\, ,~\, and
 Pran Nath\footnote{p.nath@neu.edu}
\\~\\
Department of Physics, Northeastern University,
Boston, MA 02115-5000, USA
}

\title{\textbf{An Ultralight Axion in Supersymmetry and Strings \\
and Cosmology at Small Scales}
}

\begin{document}
\maketitle
\thispagestyle{empty}

Dynamical mechanisms to generate an ultralight axion of mass $\sim 10^{-21}-10^{-22}\,\eV$
in supergravity and strings are discussed. An ultralight particle of this mass provides a candidate 
for dark matter that may play a role for cosmology at scales  $10 {\rm kpc}$ or less. {An effective
operator approach for the axion mass provides a general 
framework for models of ultralight axions, and in one case
recovers the scale $10^{-21}-10^{-22}\,\eV$ as the electroweak
scale times the square of the hierarchy with an $O(1)$
Wilson coefficient.} We discuss 
several classes of models {realizing this framework}
where an ultralight axion of the necessary size can be generated. 
In one class of supersymmetric models an ultralight axion is generated by instanton like effects. 
In the second class higher dimensional operators involving couplings of Higgs, standard model singlets,
 and axion fields 
naturally lead to an ultralight  axion.
Further, for  the class of models
considered  the hierarchy between the ultralight scale and the weak scale is maintained.  We also discuss the generation of 
an ultralight scale within string based models.  Here it is shown that in the {single modulus} KKLT moduli stabilization scheme an ultralight axion 
would require an  ultra-low weak scale. However, within the Large Volume Scenario,  
the desired hierarchy between the axion scale and the weak scale 
is achieved.
A general analysis of couplings of Higgs fields 
to instantons within the string framework is discussed and it is  shown that the condition  necessary for achieving such couplings 
 is the existence of vector-like zero modes of the instanton. 
Some of the phenomenological aspects of these models are also discussed. 

\newpage 
\section{Introduction\label{sec1}}
Recently it has been proposed~\cite{Hui:2016ltb,Marsh:2015xka} that an ultralight boson candidate for dark matter (sometimes referred to as fuzzy dark matter (FDM)), with mass
 of $\mathcal{O}(10^{-22})\, \eV$, can properly explain cosmology at scales of 10 kpc or less\footnote{Alternative possibilities for cosmology at small scales include complex dynamics or baryonic physics. However, in this work we focus on the approach involving an ultralight boson.}.
Such an ultralight particle was identified with an axion\footnote{
 For early work on axions see~\cite{Peccei:1977ur,Dine:1981rt,Weinberg:1977ma,Wilczek:1977pj,Shifman:1979if,Kim:1979if,Kim:1983dt}.} with a decay constant in the range $10^{16} \leq F \leq 10^{18}\, \GeV$. It was shown that an axion of the size needed could be generated via instanton effects.  See \cite{Gupta:2015xok}-\cite{Sarkar:2017vls} for recent works related to ultralight axions.
 
We emphasize, as did \cite{Hui:2016ltb}, that this ultralight axion is not the QCD axion. In the latter case the axion mass $m_a\simeq \Lambda_{QCD}^2/F$ depends on one parameter,
since $\Lambda_{QCD}\simeq 200$ MeV is known. For relic QCD axions produced by misalignment, this sets an upper bound $F\lesssim \,10^{12}\, \GeV$. The axion considered here is another
axion, perhaps a string axion (see e.g. \cite{Svrcek:2006yi}), that is not necessarily related to gauge dynamics in any way. Instead, its effective $\Lambda$ is set by non-perturbative
effects, such as string instantons, and therefore $m_a\simeq \Lambda^2/F$ depends on two parameters. This allows for greater freedom in the axion mass and relic abundance, and such axions are ubiquitous in string theory \cite{Arvanitaki:2009fg}.

 In this work we discuss explicit models where an {ultralight} axion can arise. 
 We will study the axion mass scale using effective operators
 and will account for the scale $O(10^{-22})\, \eV$ in terms of
 the electroweak scale and the hierarchy. We will also exhibit the emergence of such a light particle both in supergravity effective field theory
  and then in the framework of a specific class of string-motivated models.
  The outline of the paper is as follows: In Section~\ref{ORG} we discuss the general issue of the mass
  scale of the axion using an effective operator approach.  In Section~{\ref{SUSY}} we discuss  field theoretic 
 models based on supersymmetry and supergravity that lead to an ultralight axion. 
 In Section~{\ref{STRING}} we discuss the possibility of 
 realizing the axion within a more general string frameworks. Specifically we consider the KKLT and Large Volume Scenario (LVS) moduli stabilization schemes, and show that the desired hierarchy between the axion scale and the weak scale can be achieved in the latter but not the former.
  In Section~{\ref{FORMAL}}  we discuss conditions within
 string theory that allow the possibility of coupling axions
 with higher dimensional Higgs operators via D-brane ≠instantons. 
  Phenomenology of these models is discussed in Section~{\ref{PHENO}} and we conclude in Section~{\ref{CONC}}.

\section{The mass scale of the ultralight axion\label{ORG}}

An apparent conspiracy of scales exists \cite{Hui:2016ltb} between the observed dark matter relic abundance,
astrophysical observations, and common properties of axions in string theory.
Specifically, if one considers an axion in string theory with string 
scale decay constant $O(10^{16})$ GeV and demands that misalignment produces
an axion relic abundance matching the observed dark matter relic abundance $\Omega h^2 = .12$,
then the axion must be ultralight with mass $m_a$ of $O(10^{-22})\, \eV$.
This is the relevant mass scale for accounting for a variety of astrophysical observations,
as discussed in \cite{Hui:2016ltb}. 

From an ultraviolet perspective, however, it is preferable to turn this logic around:
if the mass scale $m_a\simeq 10^{-22}\eV$ could be motivated by theoretical considerations,
then a misalignment-produced axion, with string scale decay constant, would give a derivation
of the observed relic abundance. In \cite{Hui:2016ltb}
this was achieved by tuning an instanton action to obtain the
mass, which is possible in string theory but depends
critically on moduli stabilization. In this section we will
instead study axion masses utilizing symmetry arguments
and effective field theory, motivating $m_a\simeq 10^{-22} \,\eV$.

The effective operator $V_{a}$ in the scalar potential that gives the axion its mass 
must respect all of the symmetries of the theory. In particular, the axion itself
has a perturbative continuous shift symmetry that is expected to be (and typically
is in concrete constructions) broken to a discrete shift symmetry by instantons. This 
consideration leads to $V_{a} \sim \cos(a/F)$. The coefficient of this periodic term
must also respect all symmetries of the theory. Since any theory consistent with
observations respects at least standard model gauge invariance, it is natural to
decompose $V_a$ as
\begin{equation}
V_a = \tilde A \, \cO_H \, \cO_V \, \cos(a/F)\, ,
\label{OHOV}
\end{equation}
where $\cO_H$ is a hidden sector operator, and $\cO_V$ is a visible sector operator
that contains only standard model (or MSSM) fields {or a standard
model singlet $s$ that couple to the Higgs}. For this term to give the axion
a mass, both coefficient operators must receive vacuum expectation values (VEVs), where one or both could be the identity operator. Defining 
$A:=\tilde A \langle \cO_H \rangle$ and recognizing that if $\cO_V$ obtains a VEV
it can\footnote{In the MSSM we could use $(h_{u,d}^\dagger h_{u,d})^k$ and
similar conclusions would hold.}  only involve powers of $s$ and $(h^\dagger h)$, we write\footnote{In supergravity and strings
with strong dynamics  a fermion condensate of appropriate power could replace the 
$s^{2m}(h^\dagger h)^{2k}$ factor in Equation~\ref{OHOV.1}.}
\begin{equation}
V_{a} = A\,\, \frac{s^{2m}(h^\dagger h)^{2k}}{\Lambda^{4k+2m-4}} \,\, \cos(a/F)\, .
\label{OHOV.1}
\end{equation}
We note that in supersymmetric formulations higher dimensional operators with integer powers
in the superpotential will naturally lead to even integer powered higher dimensional operators. For this 
reason we take the powers of $s$ and of ($h^\dagger h)$ to be $(2m,2k)$ where $(m,k)$  
are integer or half-integer.

{ Equation~\ref{OHOV.1}} gives rise to an axion mass 
\begin{equation}\label{axmassgen}
m_a = A^\frac12 \langle h \rangle \left(\frac{\langle s \rangle}{\langle h\rangle}\right)^m\left(\frac{\langle h \rangle}{\Lambda}\right)^{n-1}\left(\frac{\Lambda}{F}\right)\, ,
\end{equation}
where $n=2k+m$ and
where $\Lambda$ is some ultraviolet cutoff. The precise axion mass depends on model dependent details that determine
the precise values of $A$, $F$, and $\Lambda$, but if $A$ is not too small and $F$ is near the high scale cutoff,
as motivated by string theory, and also
$\langle s \rangle\sim   \langle h \rangle\sim \Lambda_{\rm EW}$, we have the approximate mass equation
\begin{equation}
m_a \simeq \Lambda_{EW}\,\,\left(\frac{\Lambda_{EW}}{\Lambda}\right)^{n-1}.
\label{scales.1}
\end{equation}
For high scale cutoff $\Lambda\simeq 10^{18}\,\GeV$ and $\Lambda_{EW}\simeq 10^2 \,\GeV$,
this gives
\begin{align}
m_a & \simeq 10^{27}\, \eV \qquad \,\,\, \text{for $n=0$}\,, \nonumber  \\
m_a &\simeq 10^{11}\,\eV\qquad\,\,\, \text{for $n=1$}\,, \nonumber \\
m_a &\simeq 10^{-5}\,\eV\qquad\,\, \text{for $n=2$}\,, \nonumber \\
m_a &\simeq 10^{-21}\,\eV\qquad \text{for $n=3$}\,,
\label{scales.2}
\end{align}
and we therefore have four different regimes for axion masses: high scale, electroweak scale,
neutrino scale, and ultralight scale. Note that the mass scale relevant for ultralight
axion dark matter has arisen out of known mass scales in nature. 
 In Section~\ref{SUSY} we will show that a potential of the form 
Equation~\ref{OHOV.1} arises naturally from a superpotential, {in which
case the appearance of the singlet is related to having integral
powers of superfields.}

Concrete  analyses of some the  possibilities discussed above
for various values of $A$, $F$, and $\Lambda$ will be presented in Section~{\ref{SUSY}},
but we would like to make some brief comments here.
One critical aspect of the Section~{\ref{PHENO}} analysis will address the fact that $A$ in string theory
is typically exponentially suppressed by the volume of an internal cycle in a Calabi-Yau
manifold. From this perspective, \cite{Hui:2016ltb} used the $n=0$ case and fine-tuned this
exponential to obtain the  axion mass $O(10^{-22})\, \eV$. This requires a large internal
cycle and depends on moduli stabilization. We are simply proposing that the same small scale
can be obtained by trading instanton suppression for the electroweak hierarchy. In particular,
we will see that reasonable values of $A$ in string theory can be accommodated in this 
framework. 
In Section~\ref{FORMAL} we will discuss how operators 
of the schematic form Equation~\ref{OHOV} may arise from
D-brane instanton corrections to the superpotential
in which vector-like instanton zero-modes play
a crucial role.

\section{The axion in supersymmetry and supergravity models \label{SUSY}}

In this section we  construct explicit supersymmetric models that
generate an ultralight axion. 
The ultralight nature of the axion is due to a perturbatively-exact shift symmetry which is broken by a small amount (relative to other scales in the model) by non-perturbative effects such as instantons.  Construction of a superpotential at the perturbative level that
respects invariance under a $U(1)$ shift symmetry  $S\to e^{i\lambda} S$ for a field $S$ 
can be achieved with extra matter charged under the {standard model and} $U(1)$,  and in this case terms in the superpotential
involving $S$ and the extra matter can be written such that the superpotential is neutral
under the shift symmetry~\cite{Kim:1979if,Shifman:1979if}. Alternately
one may make the MSSM fields charged under  $U(1)$ and introduce terms
in the superpotential involving $S$ and the MSSM
fields~\cite{Dine:1981rt}. 

Here we take an alternative approach where we
introduce two fields $S_1$ and $S_2$ which are $SU(3)\times
SU(2)_L\times U(1)_Y$ singlets but  are oppositely charged under the
global $U(1)$ symmetry, i.e., under a global $U(1)$ transformation one
has  
  \begin{align}  S_1\to  e^{i\lambda} S_1, ~S_2\to  e^{-i\lambda}
S_2\,,    \end{align}  
 so that  $S_1 S_2$ is neutral under the $U(1)$.
 We consider a superpotential of the form
 \begin{align}
W_s= \mu_0 S_1 S_2 + \frac{\lambda_s}{2 M} (S_1S_2)^2\,.
\label{wslambda-s}
\end{align}
 The superfields $S_i$ (i=1,2) have the expansion
 \beqn \label{supS}
S_i
&=&\phi_i + \theta\xi_i 
+~ \theta\theta F_i \,,
\eeqn
where $\phi_i$ is a complex scalar containing the axion and the saxion, $\xi_i$ is the axino and $F_i$  the auxiliary field. Here we write
\begin{equation}
\phi_i = (\rho_i^0+ \rho_i) e^{ia_i/\rho_i^0}, ~i=1,2\, ,
\end{equation}
where the $\rho_i$ are expansions about the VEVs $\rho_i^0$. The higher dimensional operator in Equation~\ref{wslambda-s} is  needed to  give  
a VEV to  the scalar component of $\phi_i$.  The F-term equations of motion give the constraint\footnote{We will assume throughout this section that the axions are stabilized at zero, which we will find to be a consistent assumption.}
 \begin{align}
\mu_0  +  \left(\frac{\lambda_{s}}{M} \right) (\rho^0_1 \rho^0_2)=0\,.
\label{ssb}
\end{align} 
Further one finds $F\equiv \rho_1^0=\rho_2^0$. Thus we may write 
$\phi_i$ in the form 
 \begin{align}
\phi_i = (F+ \rho_i) e^{ia_i/F}, ~i=1,2\, . 
\end{align}
It is useful to define the combination of axion fields $a_1$ and $a_2$ so that 
\begin{align}
a_{\pm} = \frac{1}{\sqrt 2} (a_1 \pm a_2)\,.
\end{align}
Here one finds that Equation~\ref{wslambda-s} leads to the following potential for $a_+$
\begin{align}
V    &= 4 F^2 \mu_0^2 \left[1-\cos\left(\frac{\sqrt 2 a_+}{F}\right) \right]\,.
\label{a+mass}
    \end{align}
Equation~{\ref{a+mass} gives  $a_+$ a  mass  
   $m_{a_+}= 2\sqrt 2 \mu_0$.  One may also check that the saxion field $\rho_+$ defined so
     that  $\rho_{\pm} = (\rho_1 \pm \rho_2)/\sqrt 2$ and the axino fields 
     $\xi_{_+}$ where $\xi_{_+}= (\xi_1\pm \xi_2)/\sqrt 2$ also have exactly the same mass. 
     Thus the superpotential in Equation~\ref{wslambda-s} gives rise to an entire massive chiral multiplet 
     $\rho_+, a_+, \xi_+$, {as required by supersymmetry}.  We also note that the axion $a_-$ still possesses a      continuous shift symmetry,
     and thus no potential is generated for $a_-$ and so it remain massless. The same applies to 
     $\rho_-$ and $\xi_-$. Thus one  combination of the original chiral fields become massive 
     while the orthogonal combination remains massless.
We now turn to generation of a mass for $ a_-$. To give $a_-$ mass we need to include 
 contributions  in the superpotential which break the continuous shift symmetry. We will discuss two classes of models.
 For one class  we will use an instanton type contribution and for the other class we will use higher
 dimensional operators, which couple the Higgs fields and {standard model singlets}
  to the axion fields, which breaks the continuous shift symmetry.

 We begin by considering {models of} the first type. Here we take a superpotential  of the form
\begin{align}
W&= W_s+ W_n\,,\nonumber\\
W_n&=  A (e^{-\alpha S_1} +  e^{-\alpha S_2})\,,
\label{w-A1-A2}
\end{align}
where $W_s$ is as defined by  Equation~\ref{wslambda-s}} and 
$W_n$ violates the shift symmetry.
In this case the equations of motion  give
 \begin{align}
 \mu_0 F + (\frac{\lambda_s}{M} )F^3 - \alpha A e^{- \alpha F} =0\,.
 \label{min-condition}
  \end{align}

Retaining only the dependence on  $a_-$ 
the  axion potential takes the form
\begin{align}
V(a_-)=  2\alpha^2 A^2 e^{-2\alpha F} e^{-\alpha F \cos(a_-/\sqrt 2 F)}
[1-\cos(\alpha F \sin (a_-/\sqrt 2F))]\, .
\end{align}
We note that the form of the axion potential is not of the standard  $\cos(c a)$. However,
it reduces to it  when we expand  $\sin (a_-/\sqrt 2F)$ about $a_- = 0$ and retain the first term in the expansion.
Thus an expansion of the potential, and using the condition $\alpha F\gg1$, is needed to simulate an instanton-like
effect and leads to a mass term for $a_-$ of the form
\begin{align}
m_{a_-}\simeq   \alpha^2 A e^{-\alpha F}\,.
\end{align}
Using numbers consistent with  \cite{Hui:2016ltb}, i.e.,  
$F= 10^{17}\, {\rm GeV}, ~\alpha^2 A= 10^{12}\, {\rm GeV}$, $\alpha F = 99$, one finds
$m_{a_-}= 10^{-21} \,{\rm eV}$. A similar analysis holds for the saxion $\rho_-$ and the axino $\xi_-$ 
which develop a mass of similar size.
{We assume that $\mu_0$ is electroweak scale. Since $F=10^{17}$ GeV, this requires $\lambda_s$ to 
  be $O(10^{-12})$.\footnote{{This choice of $\lambda_s$, though small, is protected
  from renormalization by supersymmetry. We note also that this size of $\lambda_s$ can be generated in string perturbation theory; e.g. in type IIA disc instantons can generate suppressions of the form $e^{-A}$, where $A$ is the disc area. This effect 
  is distinct from the Euclidean D-brane instantons that we consider elsewhere. } }}

 Next we discuss the case when the shift symmetry is broken by a higher dimensional operator involving 
 couplings to the Higgs, standard model singlets, and the axion fields.  
 As  an organizing principle we 
  consider supersymmetric models with three sectors: visible, hidden and an overlap sector
 between the hidden and the visible sectors with interactions suppressed by Planck mass\footnote{Supersymmetric  models of this sort with three sectors have been considered 
 in previous works, see, e.g, \cite{Nath:1996qs}.} so that 
 \begin{align}
 W= W_{vis} + W_{hid}+ W_{vh} \, ,
 \end{align}
where $W_{vis}$ contains fields in the visible sector, $W_{hid}$ contains fields in the hidden sector and 
$W_{vih}$ contains the overlap.
In this analysis we assume that $W_{vis}$ contains the fields $H_1, H_2$, and $S$, where $S$ is a 
standard model singlet like the one used in the nMSSM and does not possess any shift symmetry and  $W_{hid}$ contains the axion fields $S_1,S_2$ 
discussed above. Here we take
\begin{align}
W_{vis}&=\mu_{s} S^2 + \lambda_0 SH_1H_2\, ,\nonumber\\
W_{hid}&= \mu_0 S_1S_2 + \frac{\lambda_s}{2M} (S_1S_2)^2\, ,\nonumber\\
W_{vh}&= \frac{\lambda}{M} S_1S_2 H_1 H_2 +  \frac{c}{M^{n-2}}(S_1+S_2) S^n\, .
\label{vhe}
\end{align}
We assume that the Higgs fields develop VEVs due to  sources  in the visible sector not considered here. The effects 
of $W_{vh}$ on the VEVs of $S, S_1,S_2$ are small because of Planck mass suppression. 
Thus, to the lowest order, one can see that the minimization condition in the S sector gives 
$\langle S\rangle  \sim \lambda_0 \langle v_1v_2\rangle /\mu_s$. We assume $\mu_s$ to be  electroweak size
which implies $v_0\equiv \langle S \rangle$ is electroweak size.

Next we focus on the F-term equations in the $S_1$ and $S_2$ sectors. Here we find
\begin{align}
&\mu_0 \rho_1^0 + \frac{\lambda_s}{M} (\rho^0_1)^2\rho^0_2  + \frac{\lambda}{M}  \rho_1^0 v_1v_2 
+  \frac{c}{M^{n-2}} v_0^n =0\,, \nonumber\\
&\mu_0 \rho_2^0 + \frac{\lambda_s}{M} \rho^0_1(\rho^0_2)^2  + \frac{\lambda}{M}  \rho_2^0 v_1v_2 
+  \frac{c}{M^{n-2}} v_0^n =0\,, 
\label{vhe.1}
\end{align}
From Equation~\ref{vhe.1}
we deduce $F=\rho_1^0=\rho_2^0$, which results in the constraint 
\begin{align}
\mu_0^2 F+ \frac{\lambda_s}{M} F^3  + \frac{\lambda}{M}  F v_1v_2 
+  \frac{c}{M^{n-2}} v_0^n =0\,.
\end{align}
The axion potential results from the term $\sum_{i=1,2} |\partial W/\partial {S_i}|^2$. 
Retaining only the dependence on $a_-$ we find 
\begin{align}
V(a_-) =  4c^2  \left(\frac{v_0^n}{M^{n-2}} \right)^2 \left(1- \cos(\frac{a_-}{\sqrt 2 F})\right)\, .
\label{vhe.3}
\end{align}
Equation~\ref{vhe.3} leads to mass for $a_-$ of the form
\begin{align}
m_{a_-}&=  \sqrt 2 c \frac{v_0^n}{FM^{n-2}} \nonumber\\
&= \Lambda_{\rm EW} (\frac{ \Lambda_{\rm EW}}{\Lambda})^{n-1}\, ,
\label{susy-axion.1}
\end{align}
where $v_0\sim \Lambda_{\rm EW}$, $\Lambda = (FM^{n-2})^{1/n-1}$. \\

We now show that  the term  $|\partial W/\partial {S}|^2$ does not contribute to the $a_-$ mass. The  $S$ dependent terms in the superpotential are given by 
\begin{align}
W(S)&=\mu_{s} S^2+ \lambda_0 SH_1H_2 +  \frac{c}{M^{n-2}}(S_1+S_2) S^n \, .
\label{S.1}
\end{align}
The F-term equation in this sector reads
\begin{align}
  2\mu_s S_0 + \lambda_0 v_1v_2 + 
\frac{nc}{M^{n-2}}(\rho^0_1+\rho^0_2) S_0^{n-1}  =0\, .
\label{s.3}
\end{align}
Using the result deduced above  that $\rho_a^0= F= \rho_2^0$, the axion potential from this 
this sector is given by 
\begin{align}
V_S(a_1, a_2)= |2\mu_0 S_0 + \lambda_0 v_1v_2 + 
\frac{nc}{M^{n-2}}F(e^{ia_1/F}+e^{ia_2/F}) S_0^{n-1}|^2\, .
\label{s.4}
\end{align}
Applying Equation~\ref{s.3} in Equation~\ref{s.4} we have  
\begin{align}
V_S(a_1, a_2)= |
\frac{nc}{M^{n-2}}F(e^{ia_1/F}-1+e^{ia_2/F}-1) S_0^{n-1}|^2 \, .
\end{align}
From the above we deduce that $a_-$-dependent part of the potential is 
\begin{align}
V_S(a_1, a_2)&= |
\frac{nc S_0^{n-1}}{M^{n-2}}F|^2 \left[  2 cos(\sqrt 2 a_-/F) - 8  cos(a_-/\sqrt 2F)\right]\, ,
\end{align}
which gives a vanishing mass for $a_-$.
Therefore $|F_S|^2$ does not contribute to the mass of $a_-$.
Finally we consider the potential for $a_-$ generated by the terms $\sum_{i=1,2} |\frac{\partial W}
 {\partial H_i}|^2$. Here we find 
 \begin{align}
 V_S(a_-)= \sum_{i=1,2} |\lambda_0 SH_i
+ \frac{\lambda}{M} S_1S_2 H_i|^2  \, ,
 \end{align}
 which gives a vanishing contribution to $V(a_-)$.
Superpotentials of the type
considered in $W_{vh}$ in Equation~\ref{vhe} can be generated in string models as discussed in section \ref{FORMAL}.\\

When supersymmetry is promoted to supergravity~\cite{Chamseddine:1982jx,book} and  supersymmetry 
breaking is taken into account, one will generate  soft terms and the potential will have the form 
\begin{align}
V=\sum_{i} |\frac{\partial W}{\partial \phi_i}|^2 + V_{\rm soft}\, ,
\label{soft1.1}
\end{align}
  where $\phi_i$ are all the fields that enter in the superpotential and $V_{\rm soft}$ are terms
  such as $m_0^2 \sum_i \phi_i\phi^\dagger$ and trilinear terms. In this case one finds that the
  dominant term that contributes to the axion $a_-$ mass is 
    \begin{align}
  m^2_{a_-} =  q\, h^2 \left( \frac{h}{\Lambda} \right)^{n-1}\, ,
  \end{align}
where $q$ is an $\mathcal{O}(1)$ number, and we assume $\mu_0 \sim s \sim h$. Taking $h \sim \Lambda_{EW}$, we then have 
 \begin{align}
m_{a_-}\simeq \Lambda_{\rm EW}  \left(\frac{\Lambda_{EW}}{\Lambda}\right)^{m-1} \, ,
\label{eq-5}
\end{align}
where $m  = (n +1)/2$. Here $m=3$ requires $n=5$.

\subsection{Models with higher dimensional Higgs-axion couplings}

Next we discuss the case when the shift symmetry is broken by a higher dimensional operator involving 
 couplings of the Higgs and $S_i$. Here we assume a superpotential of the form 
 \begin{align}
W&=  \mu_0 S_1S_2 + \frac{\lambda_s}{2M} (S_1S_2)^2 + \frac{\lambda}{M}S_1S_2 H_1H_2
+  \frac{c}{M^{2k-2}} (S_1+ S_2) (H_1H_2)^k\,.
  \label{s4h}
  \end{align}

Next using the superpotential  of Equation~\ref{s4h} and 
after spontaneous breaking which gives VEVs to  $S_i$ and also assuming that $H_i$
develop VEVS,  axion $a_-$  potential can be obtained as discussed in the previous analysis and one gets

   \begin{align}
   V(a_-)  &=  \left[(\frac{2}{M^{2k-2}}  c (v_1v_2)^k)^2  
    +   
  (\frac{cF}{M^{2k-2}}  (v_1v_2)^{k-1})^2  (v_1^2+v_2^2)\right]
(1-\cos(\frac{a_-}{\sqrt 2 F}))   
\label{pot.2}
      \end{align}
 For the case $k=2$ the first term in the brace on the right hand side of  Equation~\ref{pot.2} is small relative to the second
 which gives an axion mass 

\begin{align}
M_{a_-}= c (v_1^2+ v_2^2)^{1/2}  \left(\frac{M}{F}\right)\left(\frac{(v_1v_2)^{1/2}}{M}\right)^{2k-2}
\end{align}
   This is of the form  Equation~\ref{scales.1} with $n=2k-1$ and for $k=2$ one has $n=3$ which
   gives the ultralight axion. We note that after soft terms are taken into account we will
   have a result similar to Eq (\ref{eq-5}).

\vspace{1cm}
As a final example we consider a model where the axion couples directly to the Higgs fields, via a non-perturbative term in the superpotential. We present this model 
because it is a very simple realization of the organizing principle of Section \ref{ORG} involving higher dimensional Higgs-axion couplings. In this example the axion $a$ is the imaginary part of a complex modulus $T = \tau + i\, a$, whose potential is generated non-perturbatively. This class of models is ubiquitous in string theory, and we will explore the details of string embeddings in Sections~\ref{STRING} and~\ref{FORMAL}. We consider a superpotential of the form

\begin{align}
W = W_0 + \mu H_1 H_2  + \Lambda^{3-2n} (H_1 H_2)^n e^{-T/F}\, ,
\end{align}

where $W_0$ is a constant obtained from integrating out heavy fields. The axion appears in the potential only via the $H_1$ and $H_2$ F-terms, and a quick calculation shows the mass of $a$ takes the form
\begin{align}
m_a = 2 \left( \frac{h}{\Lambda} \right)^n \sqrt{n\,  \mu \frac{\Lambda^3}{F^2} e^{-\tau} } \, .
\end{align}
Taking $F \sim \Lambda$ to be a high scale and $h \sim \mu \sim \Lambda_{EW}$, we have
\begin{align}
m_a = 2\sqrt{n} \left( \frac{ \Lambda_{EW}}{\Lambda} \right)^n \sqrt{ \Lambda_{EW} \Lambda e^{-\tau}} \, .
\end{align}
Furthermore, if we take $ \Lambda e^{-\tau} \sim \Lambda_{EW}$, we find
\begin{align}
m_a \simeq  \Lambda_{EW} \left( \frac{ \Lambda_{EW}}{\Lambda} \right)^n \, .
\end{align}
Here taking $n=2$ provides the desired ultralight mass for the axion. In many string models \cite{Blumenhagen:2006xt,Florea:2006si,Ibanez:2006da} the $\mu$-term in the superpotential is generated non-perturbatively, so we find it plausible that additional non-perturbative effects could generate this coupling at the same scale. Alternatively, it may be possible for the instanton that generates the higher order Higgs coupling to be in the same homology class as the instanton that generates the $\mu$-term; in this case the relationship $\Lambda e^{-\tau} \sim \Lambda_{EW}$ is automatic. We leave the study of these important global issues to future work.

\section{Axions in simplified string models \label{STRING} }
The authors of~\cite{Hui:2016ltb} suggest that the FDM model of dark matter could be embedded in a string compactification, and the necessary mass and axion decay constant are natural from a stringy point of view. To make a precise statement one should  scan over an ensemble of vacua and use the distribution of axion masses and decay constants to estimate the frequency in which parameters consistent with FDM occur. Unfortunately, while it is well-known how to calculate axion decay constants even when the number of moduli is large (c.f.~\cite{Long:2014fba}), calculating the masses requires intimate knowledge of non-perturbative effects, which are currently only partially calculable. In addition, moduli stabilization with a large number of moduli is notoriously difficult. 

It is therefore our goal to find a realistic simplified model to demonstrate that embedding FDM in string theory is consistent with moduli stabilization, and does not remove us from the regime of validity of the effective theory. 

A typical 4d effective SUGRA theory constructed from a string compactification has scalar fields known as moduli. These fields arise from reducing the metric and various $p$-form gauge fields along appropriate $p$-cycles in the internal space $X$. A virtually universal class of moduli are the K\"ahler moduli, whose vacuum expectation values parameterize complexified volumes of holomorphic cycles in $X$. We consider a compactification of IIB string theory on a Calabi-Yau orientifold $X$, which yields an effective $\mathcal{N} = 1$ SUGRA theory in 4d. Type IIB string theory has a four-form gauge field $C_4$ in 10d, and dimensionally reducing $C_4$ along a holomorphic four-cycle (divisor) in $X$ yields an axion in the 4d theory. This axion pairs with the volume modulus of the four-cycle in a complex scalar field, which is the lowest component of a chiral superfield. The K\"ahler moduli $T^i$ are written as
\begin{equation}
T^i = \frac{1}{2} \int\limits_{D^i} J \wedge J + i\,  \int\limits_{D^i} C_4 \equiv \tau^i + i\, \theta^i \, ,
\end{equation}
where $D^i$ is the corresponding divisor with volume modulus $\tau^i$ and axion $\theta^i$, and $J$ is the K\"ahler form on $X$.  
The theory typically has other moduli besides K\"ahler moduli, including the complex structure moduli $U$ and the holomorphic axio-dilation $S = e^{-\phi} + i\, C_0 \equiv S_1 + i\, S_2$.\footnote{Here the variables $S$, $S_1$, and $S_2$ are not to be confused with the ones from Section~\ref{SUSY}.} The tree-level K\"ahler potential takes the form
\begin{equation}
K =  - \, \text{log} (S+ \bar{S}) -2 \, \text{log} (\mathcal{V}  ) + K_{cs} (U, \bar{U})  \, .
\end{equation}
The complex structure moduli and holomorphic axio-dilaton acquire masses via the tree level flux superpotential~\cite{Gukov:1999ya}
\begin{equation}
W_{\text{Tree}} = \int_X G_3 \wedge \Omega\, ,
\end{equation}
where $G_3$ is a particular flux on $X$, and $\Omega$ is the holomorphic $(3,0)$-form. We will assume that $S$ and $U$ are stabilized at a high scale by $W_{\text{Tree}}$. The K\"ahler moduli, on the other hand, only appear in the superpotential non-perturbatively~\cite{Witten:1996bn}. Including these non-perturbative effects, the superpotential then takes the form
\begin{equation}
W = W_0 + \sum\limits_{a} A_a e^{-q^{a}_{\, i} T^i}\, ,
\end{equation}
where $W_0 = \langle W_{\text{Tree}} \rangle$, and the matrix $q^{a}_{\, i}$ is a matrix of rational numbers.

 \subsection{KKLT moduli stabilization \label{KKLT}}
In this section we discuss the KKLT moduli stabilization scheme~\cite{Kachru:2003aw}, in which the classical superpotential is balanced against an exponentially small non-perturbative effect in order to stabilize the K\"ahler moduli.  We wish to see if an ultralight axion can be generated within the KKLT scheme. We consider the $D=4,\, \mathcal{N}=1$ supergravity (SUGRA) potential~\cite{Chamseddine:1982jx,book}:
\begin{align}
& V = e^{\kappa^2 K} (K^{i\bar j} D_iW D_{\bar j} \bar W - 3\kappa^2 |W|^2)\,,\nonumber\\
& D_iW=  W_{,i} + \kappa^2 K_{,i} W\,,
\label{vsug}
\end{align}
In the analysis below we set $\kappa = 1$. In the case of a single \K modulus the \K potential can be written as\footnote{In this section we suppress the dependence on $S$ and $U$ as they will only contribute an overall scale.}
\begin{align}
K = -3 \, \text{log}(T+ \bar T)\,,
\end{align}
and the superpotential takes the form 
\begin{align}
W= W_0 +A  e^{- qT}\,,
\label{w0A}
\end{align}
where $A$ and $W_0$ is independent of $T$. Without loss of generality we  assume $A$ and $W_0$ are real.  Let us now expand  $V$ in the following form 
\begin{align}
V=  e^K \left(K^{T\bar T} \partial_TW \partial_{\bar T} \bar W 
+K^{T\bar T}
 (\partial_TK   W \partial_{\bar T} \bar W + 
  \partial_{\bar  T} K  \bar W \partial_{T}  W) \right)\,.
 \end{align}
Using the decomposition 
\begin{align}
T= \tau+ i\,\theta\,,
\end{align}
 $V$ takes the form 
\begin{align}
V= \frac{1}{6\tau}
\left[
 q^2A^2  e^{- 2q\tau}
  +\frac{3 qA W_0}{\tau} e^{- q \tau } \cos(q\,\theta)
  +  \frac{3 qA^2}{\tau} e^{-2q \tau}  \right]\,.
 \end{align}
Solving the F-term equations $DW = 0$ one finds\footnote{We note that $DW=0$ gives exactly
the same condition for the critical point as the minimization of the  potential in this case.}
\begin{equation}\label{eqn:crit}
W_0 = -A e^{-q\tau_0} \left(1 + \frac{2}{3} q \tau_0\right)\, ,
\end{equation}
where $\tau_0 = \langle \tau\rangle$. 
We expand around the critical point so that 
$\tau= \tau_0 + \tau',  \langle \theta\rangle=0$.
The kinetic energy then takes the form 
\begin{align}
L_{kin}= - \frac{3}{4\tau_0^2} 
\left[\partial_\mu \tau' \partial^\mu \tau' + \partial_\mu \theta \partial^\mu \theta\right]\,.
\end{align}
We define the canonically-normalized fields
\begin{align}
\rho \equiv \frac{\sqrt 3}{\sqrt 2 \tau_0} \tau',~ a \equiv  \frac{\sqrt 3}{\sqrt 2 \tau_0}\theta\,,
\end{align}
for which the kinetic energy takes the canonical  form. We have 
\begin{align}
~~V(a)&=  \delta (1- \cos(\gamma a))\,,\nonumber\\
\delta = &-\frac{qAW_0}{2 \tau_0^2}e^{-q\tau_0},
\gamma = \frac{\sqrt 2 q \tau_0}{\sqrt 3}\,.
\end{align}
At the AdS minimum, the mass of $a$ can be written as
\begin{equation}
m_a = \frac{1}{3}A e^{- q \tau} q^{3/2} \sqrt{3 + 2 q \tau}\, ,
\end{equation}
where we have used Equation~\ref{eqn:crit} to evaluate the mass at the minimum of the potential. 
In gravity mediated breaking of supersymmetry (see \cite{book} and the references therein)
the weak scale  $m_s$  is related to the hidden sector $W_0$ so that 
  $m_s=e^{K/2} |W_0|$ .
Setting $q=2\pi, A=1$ and stabilizing the modulus $T$ one finds that an axion mass of 
$10^{-22}\, \eV$ requires {the string scale to be far below the electroweak scale.}

 {Thus, we see that single modulus
KKLT is incompatible with an ultralight axion.}

\subsection{The Large Volume Scenario}

In Section~\ref{KKLT} we found that the single modulus KKLT realization of FDM had a separation of scales issue. In order to get around this we must modify the theory, by introducing more fields and/or by considering further corrections to the potential. A particularly simple way to introduce an additional scale is to consider the first non-vanishing $\alpha^{\prime}$-correction to the K\"ahler potential. This correction was computed in~\cite{Becker:2002nn}, and the corrected K\"ahler potential takes the form\footnote{In this note we work exclusively in the Einstein frame.}  
\begin{equation}
K =  - \, \text{log} (S+ \bar{S}) -2 \, \text{log} (\mathcal{V}  +\alpha) 
+ K_{cs} (U, \bar{U})  \, ,
\end{equation} 
where $\alpha = \frac{1}{2} \xi S_1^{3/2}$, $\xi = \zeta (3) \chi/2 (2\pi)^3$, and $\chi$ is the topological Euler characteristic of $X$. 
The \textit{Large Volume Scenario} (LVS)~\cite{Balasubramanian:2005zx} is a multi-modulus ($\geq 2$) stabilization scheme that uses the $\alpha^{\prime}$-correction, along with a non-perturbative effect, to realize a hierarchy of scales.  

 Here we will consider the simplest case, where the number of K\"ahler moduli, which is counted by the Hodge number $h^{1,1}(X)$, equals two. It was shown in~\cite{Gray:2012jy} that the volume all $h^{1,1} = 2$ Calabi-Yau manifolds can be written in the \textit{Strong Cheese} form, such that
\begin{equation}
\mathcal{V}  = \eta \,( \tau_b^{3/2} - \tau_s^{3/2}) \, .
\end{equation}
Here $\tau_b$ is the big (or large) cycle, which controls the overall volume (size of the cheese), and $\tau_s$ is a small cycle (a hole in the cheese). The constant $\eta$ is typically an $\mathcal{O}(1)$ number, which depends on the intersection numbers of $X$. We will take $\eta = 1/9\sqrt{2}$ for concreteness, as in the $\mathbb{P}^4_{1,1,1,6,9}$ Calabi-Yau hypersurface. Each of these volume moduli pairs with an axion, so we have two complex scalers $T_s = \tau_s + i \,\theta_s$ and  $T_b = \tau_b + i \,\theta_b$.

In LVS the overall volume is taken the be large, with $\tau_s$ left small, so that $V \sim  \tau_b^{3/2}$, and
\begin{equation}
\frac{ \tau_s}{\tau_b} \ll 1, \quad \frac{\alpha}{\mathcal{V}} \ll 1\, .
\end{equation}
In this regime that K\"ahler potential can be expanded as
\begin{equation}
K \approx  -2 \, \text{log} (\mathcal{V} ) - 2\,\frac{\alpha}{\mathcal{V}}\, .
\end{equation}
In standard LVS the cycle $\tau_b$ is taken to be large enough to effectively ignore any non-perturbative effects that depend on $\tau_b$. The superpotential then takes the form
\begin{equation}\label{eqn:npsmall}
W = W_0 + A_s e^{-a_s T_s}\, .
\end{equation}

The axion $\theta_b$ is massless in this approximation, as it does not appear in the potential. Of course, it is expected that a non-perturbative correction to the potential will generate a mass of $\theta_b$. In an $\mathcal{N}=1$ SUGRA model the mass for $\theta_b$ can be generated by a correction to either the superpotential or the K\"ahler potential (or both). Let us first consider a correction to the superpotential, of the form
\begin{equation}\label{eqn:nplarge}
\Delta W  = A \,e^{-a_b T_b}\, .
\end{equation}
At large volume (large $\tau_s$) this correction is negligible compared to the terms in Equation~\ref{eqn:npsmall}, and will therefore not affect the stabilization of $\tau_b$, $\tau_s$, or $\theta_s$. However, Equation~\ref{eqn:nplarge} provides the only term in $W$ that explicitly depends on $\theta_b$, and will therefore be the leading-order operator that generates a mass for $\theta_b$, in the absence of additional corrections. However, this term will be quite suppressed, and so one must consider whether this correction truly is leading order. Holomorphy, along with the shift-symmetry of the axion, constrain $\Delta W$ to take the form derived in~\cite{Witten:1996bn}:
\begin{equation}
\Delta W = \sum\limits_i A_i e^{-q^{i}_{\, j} T^j}\, ,
\end{equation}
where the $q^{i}_{\, j}$ are rational numbers, and $-q^{i}_{\, j} \tau^j$ is a positive rational multiple of the volume of a divisor.

However, holomorphy does not constrain the K\"ahler potential, and the corrections can take a more general form. It is beyond the scope of this work to explicitly calculate any such corrections; instead, we believe the following assumptions are well-motivated:
\begin{enumerate}
\item $\Delta K$ is periodic in $\theta_b$.
\item $\Delta K$ is generated by instantons that are charged under $C_4$; namely, Euclidean D3 and anti-D3 branes.
\item The nonperturbative correction preserves the logarithmic form of the K\"ahler potential.
\end{enumerate}

If one assumes that the correction is generated by Euclidean D3 or anti-D3 brane, wrapping a cycle $\gamma$, then we expect the correction to the K\"ahler potential to take the form
\begin{equation}\label{eqn:kcorr}
\Delta K = \frac{A}{\mathcal{V}} e^{-S} f(\theta_b)\, ,
\end{equation}
where $f$ is a periodic function of $\theta_b$. Here $S$ is the instanton action, which we expect to go roughly as the volume of the brane. In order to to solve the equations of motion $\gamma$ should be a \textit{locally} volume minimizing representative of its class $[\gamma ]$, with volume vol$(\gamma)$, and so $S \simeq \text{vol}(\gamma)$. However, since this instanton is correcting the K\"ahler potential, and not the superpotential, $\gamma$ does not need to have minimal volume in the class $[\gamma]$, as it is not necessarily a holomorphic representative. Therefore, $ \text{vol}(\gamma) \geq \tau_\gamma$, where $\tau_\gamma$ is the minimal volume of $[\gamma]$. Without an explicit calculation we see no reason to assume that the inequality $ \text{vol}(\gamma) \geq \tau_\gamma$ cannot be saturated by at least some corrections to the K\"ahler potential. If this is the case the correction in Equation~\ref{eqn:kcorr} could provide corrections to $V$ of the same order as those in Equation~\ref{eqn:nplarge}. We will assume this is not the case, but it is important to understand these corrections further in the future.

Under the assumption that the correction to $W$ given in Equation~\ref{eqn:nplarge} provides the leading order term for $\theta_b$, the scalar potential\footnote{We set $K_{\text{cs}} = 0$ for simplicity, and absorb any phase of $W_0$ into the axions.} takes the form
\begin{align}\label{eqn:potential}
V = &\left( \frac{12 \sqrt{2} |A_s|^2 a_s^2 \sqrt{\tau_s} e^{-2 a_s \tau_s}}{\mathcal{V} S_1} + \frac{2 |A_s W_0|a_s \tau_s e^{- a \tau_s}}{\mathcal{V}^2 S_1} \text{cos}(a_s \theta_s) \right. \nonumber \\& \left. + \frac{2 a_b \tau_b |A_b W_0|}{\mathcal{V}^2 S_1} e^{-a_b \tau_b} \text{cos}(a_b \theta_b) + \xi \frac{3 |W_0|^2 \sqrt{S_1}}{8 \mathcal{V}^3}  \right. \nonumber \\ &\left.  + \frac{4 a_b a_s \tau_b \tau_s |A_b \bar{A}_s|}{\cV^2 S_1} e^{-a_s \tau_s-a_b \tau_b} \text{cos}(a_b \theta_b - a_s \theta_s) \right) \, .
\end{align}

This form is derived in the appendix. The axions are stabilized at  $ \theta_b = \pi/a_b, \theta_s = \pi/a_s$. A non-supersymmetric AdS minimum of the potential is found approximately at $a \tau_s \sim \text{ln} \mathcal{V}$. For a concrete example we consider the following parameters:
\begin{align}\label{eqn:params}
h^{1,1} = 2, h^{2,1} = 171, W_0 = 10^{-12}, A_s = A_b = 1, a_s = a_b = 2\pi/6, S_1 = 10.71 \, .
\end{align}
These numbers are well-motivated in weakly-coupled IIB string theory. Calabi-Yau manifolds with a hierarchy  in $h^{1,1}$ and $h^{2,1}$ are quite common, and the dual Coxeter number $6$ appearing in $a_s$ and $a_b$ corresponds to an SO(8) gauge group, which is consistent with our weak coupling assumption. Here we also have $S_1 = 1/g_s$, so in this example $g_s \approx 0.1$ is small. Inserting these parameters into Equation~\ref{eqn:potential} and minimizing the potential, we find the volume\footnote{In this section we express all of our volumes in the appropriate units of $\alpha^{'}$.} is stabilized at $\mathcal{V} = 187$. The small cycle is stabilized at $\tau = 32.5$. One might be concerned that a volume of $\mathcal{O}(100)$ is too small for the $1/\mathcal{V}$ expansion of the K\"ahler potential to be valid, but in this example the correction is at the percent level, so we expect the approximation to be good\footnote{While the relative smallness of the perturbation to the K\"ahler potential is a necessary condition for the LVS approximation to be valid, it is not sufficient, due to the non-trivial K\"ahler geometry. We have checked that the higher order terms are subleading.}. Using the parameters in Equation~\ref{eqn:params} we find a light axion mass of $3.9 \times 10^{-22}$ eV. The mass of the other axion is approximately $26$ TeV, and the masses of the saxions are 590 GeV and 280 TeV. The fermions masses are 13 TeV and 26 TeV. Both axion decay constants are $\mathcal{O}(10^{16})$ GeV. Importantly, the gravitino mass, which is the order parameter for SUSY breaking, is not too large, at approximately 13 TeV. It would be difficult to argue for SUSY as a solution to the hierarchy problem if the gravitino mass was near the Planck scale.

While the potential in Equation~\ref{eqn:potential} is a toy-model for a real string compactification, with all relevant corrections computed, our analysis demonstrates the a mass scale for the lightest axion of $\mathcal{O}(10^{-22})$ eV is arguably consistent with moduli stabilization and a realistic electroweak scale. Of course, further study of both non-perturbative and perturbative corrections to the K\"ahler potential, such as those in~\cite{Berg:2007wt}, and superpotential is important in understanding how FDM could be embedded in string theory.

\section{Ultralight Axion Couplings to the Higgs in String Theory\label{FORMAL}}

In this section we discuss how operators of the
form $(h^\dagger h)^n cos(a/F)$ may arise in string theory, focusing on non-perturbative
corrections to the superpotential \cite{Witten:1996bn}.
Some of the concepts implicit in previous sections will
be repeated here in order to present a more complete 
picture of instanton corrections to the superpotential
in string theory.

Non-perturbative corrections to the superpotential may arise 
from gauge dynamics, Euclidean D-brane instantons,
M2-brane instantons, or worldsheet instantons, depending
on the situation. For example, in type IIB compactifications, in 
particular in KKLT and LVS, Euclidean D3 (ED3) instantons may generate such corrections, 
and Euclidean D2 (ED2) instantons and M2-brane instantons provide similar corrections
in type IIA and M-theory compactifications. The non-perturbative contribution to the
superpotential from a single instanton is typically written in the schematic form
\begin{equation}
W_{np} = A(\phi) e^{-T}\, ,
\end{equation}
where $T$ is a modulus appropriate to the compactification, e.g. a K\" ahler modulus
in type IIb compactifications, where $\langle \text{Re}(T) \rangle = vol(D)$, with
$D$ the internal cycle wrapped by the instanton, and the axion $a$ is $\text{Im}(T)$. $A(\phi)$ is an instanton prefactor 
that depends on other moduli. These couplings do not couple $a$ to the Higgs, and therefore are
not of the desired type.

More general classes of brane instantons exist~\cite{Blumenhagen:2006xt,Florea:2006si,Ibanez:2006da}
 in which the instanton prefactor may also contain gauge invariant combinations of chiral supermultiplets charged
under gauge groups. Such corrections arise due to the presence of additional
instanton zero modes when $D$ intersects some other cycle $D'$ 
wrapped by spacetime-filling branes that carry non-trivial gauge sectors. We write
the general form of these corrections as 
\begin{equation}
W_{np} = A(\phi)\,\, \cO_H \cO_V \, e^{-T}\, ,
\label{effectiveV}
\end{equation}
where the visible sector operator $\cO_V$ contains only MSSM superfields, whereas $\cO_H$
may have charged fields beyond the MSSM, which could live in a hidden sector separated from
the visible sector in the extra dimensions. One important aspect of these instantons is that
they may generate the leading coupling in
$\cO_H \cO_V$, if $\cO_H\cO_V$ on its own is forbidden by an anomalous $U(1)$ symmetry. For example,
in weakly-coupled type II compactifications the top-quark Yukawa Coupling $10\, 10\, 5$
of a Georgi-Glashow $SU(5)$ GUT is always forbidden in perturbation theory, as are the
flavor-diagonal Majorana mass terms for right-handed neutrinos. Obtaining these
superpotential couplings therefore \emph{requires} non-perturbative effects, such
as the ones described.

For concreteness, we will restrict our attention to ED3
instantons in type IIb compactifications, though similar
statements regarding vector-like zero modes and higher
dimensional operators should hold in other contexts as well.

\vspace{.5cm}

We would like to study situations under which an ultralight
axion mass can arise from an effective operator of the
schematic form (\ref{OHOV}), which itself arises from
an instanton contribution to the superpotential.
For this to happen, 
holomorphy and gauge invariance dictate that the
non-perturbative superpotential contains a term\footnote{One could easily incorporate the field $S$, considered in Section~\ref{ORG}, in this effect, but we omit it here for simplicity of discussion.}
\begin{equation}
\label{Wax}
W_{ax} = A\,\frac{(H_1H_2)^n}{M_s^{2n-3}}e^{-T}.
\end{equation}
Whether or not such a term exists depends on the 
detailed structure of the instanton zero modes.
These include ED3-ED3 zero modes, as well as ED3-D7
zero modes that arise
from ED3 intersections with spacetime filling D7-branes
that give rise to the Higgs fields $H_1$ and $H_2$.
Of particular important are the fermionic zero modes,
so-called $\lambda$-modes, in the ED3-D7 sector.

For example, if the $\mu$-term $H_1H_2$ is forbidden 
by an anomalous $U(1)$ symmetry, a non-perturbative
effective of the form
\begin{equation}
A M_s \, H_1 H_2 \, e^{-T}
\end{equation}
may generate it non-perturbatively~\cite{Blumenhagen:2006xt,Florea:2006si,Ibanez:2006da}, where the 
effective $\mu$ parameter $\mu_{eff} = A M_s e^{-\langle Re(T)\rangle}$ may be at the electroweak scale depending on
the expectation value of the stabilized field $T$. In
this way, ED3-instantons give a solution to the $\mu$-problem.
Generating such an operator that is forbidden in perturbation
theory by an anomalous $U(1)$ symmetry requires a chiral
excess of $\lambda$-modes and an associated shift 
of $T$ under the anomalous $U(1)$, so that the entire
operator is gauge invariant. In such a case the axion
in $T$ becomes the longitudinal component of the massive
$Z'$ boson associated to the anomalous $U(1)$, which has a string
scale mass via the St\" uckelberg mechanism. See \cite{Cvetic:2009yh,Cvetic:2009ez}
for systematic phenomenological studies in this context.

For an ultralight axion to appear in $\eqref{Wax}$,
it is necessary for it to not be eaten via
the St\" uckelberg mechanism. Therefore the
operators $(H_1H_2)^n$ must not be forbidden by
an anomalous $U(1)$, and correspondingly the instanton
must have at most vector-like $\lambda$-modes, i.e. the
modes have index zero. Using the instanton calculus of
\cite{Blumenhagen:2006xt}, an instanton on a divisor
$D$ with K\" ahler modulus $T$ and a single vector-like pair
$\lambda \overline \lambda$ with an appropriate
structure of ED3-ED3 zero modes generates an effective
operator of the form
\begin{equation}
\int d^4x d^2\theta \int d\lambda d\overline\lambda
\,\, A M_s^3 \,\, e^{-T + \lambda H_1 H_2 \overline  \lambda / M_s^2 + \dots}
\supset 
\int d^4x d^2\theta A M_s H_1 H_2 e^{-T},
\end{equation}
which is precisely $W_{ax}$ in the $n=1$ case.
More generally, there may be $n$ pairs of vector-like
zero-modes $\lambda_i \overline \lambda_i$, in
which case there are more Grassmann integrals, and we have
\begin{equation}
\int d^4x d^2\theta \int d\lambda_1 d\overline\lambda_1
\dots d\lambda_n d\overline \lambda_n
\,\, A M_s^3 \,\, e^{-T + a_{ij} \lambda_i H_1 H_2 \overline\lambda_j / M_s^2 + \dots}
\supset 
\int d^4x d^2\theta \,\, det(a_{ij}) A \frac{(H_1 H_2)^n}{M_s^{2n-3}} e^{-T},
\end{equation}
which is precisely $W_{ax}$. Thus, we see that a
superpotential operator $W_{ax}$ of the desired form
may be generated if there is an instanton with $n$
pairs of vector-like zero modes $\lambda\overline\lambda$.
The $n=2$ case is quite similar to the non-perturbative Weinberg
operator $LH_2LH_2$ studied in \cite{Cvetic:2010mm}, since $L$ and $H_1$
have the same quantum numbers under the MSSM gauge group.

The appearance $W_{ax}$, then, depends crucially on the
structure of vector-like instanton zero modes, and we
would like to consider when such zero modes exist.

Suppose that an ED3 and a D7-brane (or a stack of D7-branes) wrap divisors $D$ and $D'$ in a smooth Calabi-Yau threefold $X$
that intersect along a curve $C:= D\cdot D'$. Both the instanton and the D7-brane may carry $(1,1)$-form
worldvolume fluxes (or more generally holomorphic vector bundles), which may be
written in terms of line bundles $\cL_D$ and $\cL_{D'}$ on $D$ and $D'$, respectively.
Then the ED3-D7 instanton zero-modes at the intersection are counted by the 
cohomology $h^{i}(C,K_C^{1/2} \otimes \cL)$, where $\cL:= \cL_D|_C \otimes \cL^{-1}_{D'}|_C$.

As discussed, a necessary condition for obtaining couplings of the desired type is that there is no
chiral excess of ED3-D7 zero modes on $C$, i.e.
\begin{equation}
\chi(C,K_C^{1/2} \otimes \cL) = h^0(C,K_C^{1/2} \otimes \cL)-h^1(C,K_C^{1/2} \otimes \cL)=0\, .
\end{equation}
Computing this index by applying the Hirzebruch-Riemann-Roch theorem, we have
\begin{equation}
\chi(C,K_C^{1/2} \otimes \cL)=\int_C ch(K_C^{1/2} \otimes \cL)td(C)=\int_C (1+c_1(K_C^{1/2} \otimes \cL))(1+c_1(C)/2) = \int_C c_1(\cL)\, ,
\end{equation}
and we see the index is zero when $c_1(\cL)=0.$ 
By this we see that if
$c_1(\cL_D|_C)=c_1(\cL_{D'}|_C)$ then $\chi(C,K_C^{1/2} \otimes \cL)=0$, i.e. we have
at most vector-like instanton zero modes on $C$.

In such a case, determining whether there actually are vector-like instanton zero modes
requires computing the cohomology, not just the index. This computation can be done by
a variety of means, but as an existence proof we would like to present a simple example.

Consider the case where a divisor $D=\bP^1 \times \bP^1$ 
is wrapped by an ED3 instanton that intersects a space-time filling D7-brane on another
divisor $D'$ at a degree
$(m,n)$ curve $C\subset D$, and there are no worldvolume fluxes, i.e. $\cL_D=\cO_D$ and
$\cL_{D'}=\cO_{D'}$. The zero modes are counted by $h^{i}(C,K_C^{1/2})$, which has
index zero, where $K_C = (K_D+\cO(C))|_C = \cO(m-2,n-2)|_C$. Taking the square root,
a Koszul sequence for $K_C^{1/2}$ is given by 
\begin{equation}
\label{eqn:koszul}
0\rightarrow \cO_D\left(-\frac{m}{2}-1,-\frac{n}{2}-1\right) \rightarrow \cO_D\left(\frac{m}{2}-1,\frac{n}{2}-1\right) \rightarrow K_C^{1/2}\rightarrow 0\, .
\end{equation}
By Serre duality, $h^i(D,\cO_D(\frac{m}{2}-1,\frac{n}{2}-1)) =h^{2-i}(D,\cO_D(-\frac{m}{2}-1,-\frac{n}{2}-1))$. Since a degree
$l$ line bundle on $\bP^1$ has $l+1$ global sections,
and therefore a degree $(k-1,l-1)$ line bundle on $\bP^1\times \bP^1$ has $kl$ global sections, $h^i(D,\cO_D(\frac{m}{2}-1,\frac{n}{2}-1))=(mn/4,0,0)$. Using
the long exact sequence in cohomology associated to the Koszul sequence \eqref{eqn:koszul},
we obtain
\begin{equation}
h^i(C,K_C^{1/2}) = \left(\frac{mn}{4},\frac{mn}{4}\right),
\end{equation}
which shows that there are vector-like instanton zero modes for general even $m$ and $n$. For a more in depth introduction to this type of computation, see e.g.~\cite{Cvetic:2010rq,Cvetic:2010ky}.

\begin{figure}[t]
\hspace{-1cm}
\includegraphics[scale=1.1]{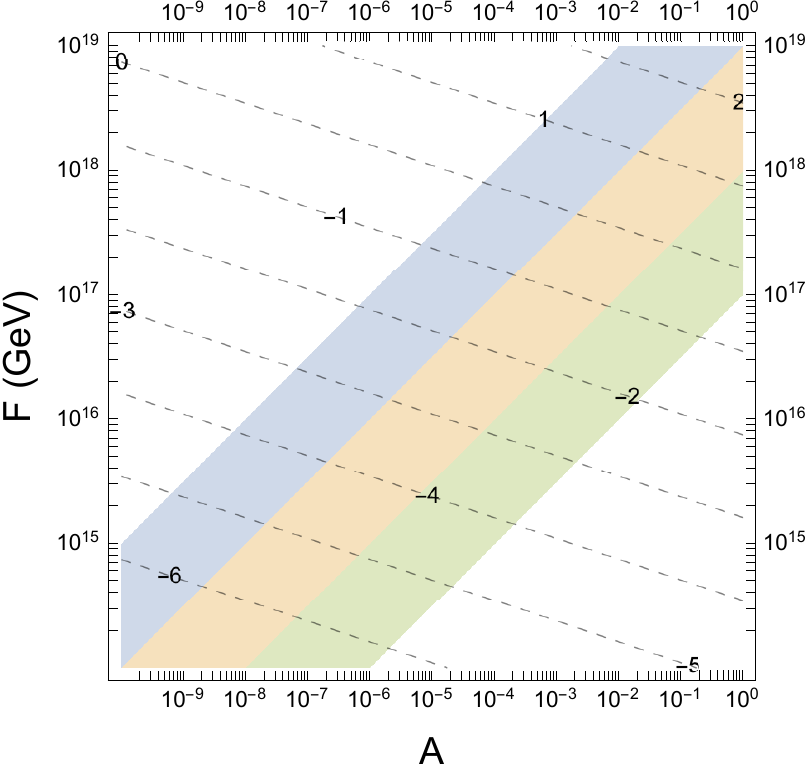}
\includegraphics[scale=1.13]{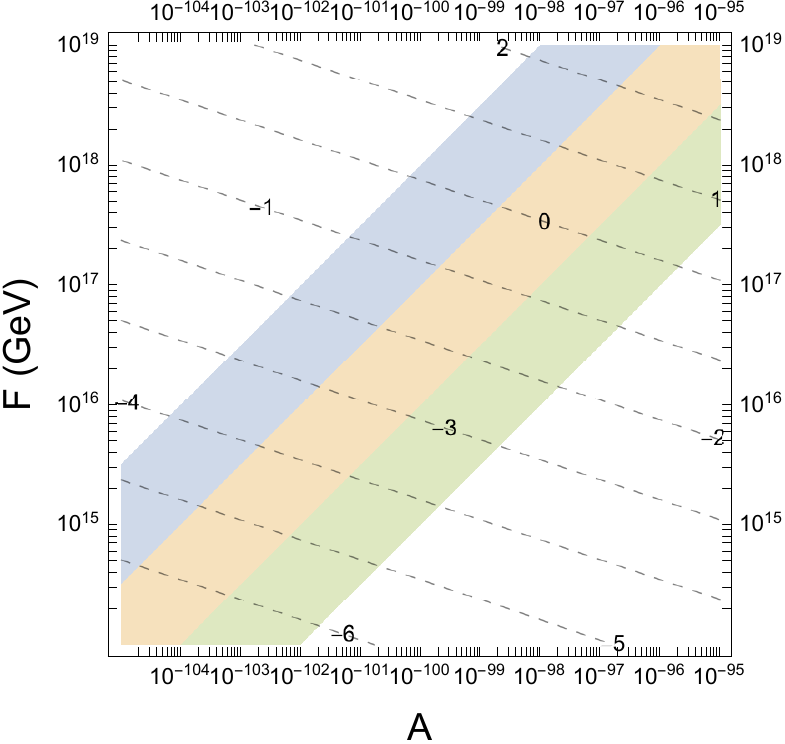}
\caption{Axion relic abundance and mass as a function of 
axion decay constant $F$ and Wilson coefficient A. 
The dashed contours denote the axion relic abundance and
are labelled by $log_{10}(\Omega_{ax}h^2)$; the $-1$ contour
is the observed relic abundance. The blue, orange, and
green bands are mass regions 
$10^{-23}\eV \leq m_a \leq 10^{-22}\eV$,
$10^{-22}\eV \leq m_a \leq 10^{-21}\eV$,
and $10^{-21}\eV\leq m_a \leq 10^{-20}\eV$, respectively,
so that the $m_a=10^{-22}\eV$ line is the boundary
between
the blue and orange bands. \emph{Left:} the $n=3$ case,
which accommodates the relic abundance and mass by using
the electroweak hierarchy. \emph{Right:} the $n=0$ case,
which accommodates these solely with instanton suppression.}
\label{figrelicmass}
\end{figure}

\section{Phenomenology\label{PHENO}}
  As discussed in~\cite{Hui:2016ltb}, the relic density of the ultralight axion arises from misalignment, where after inflation the axion begins to oscillate around its minimum. Initially the
  axion field is assumed to have a value close to the decay constant, 
  which leads to  a relic density 
\begin{align}
\Omega_{a} \sim 0.1\left(\frac{m_{a}}{10^{-22} \eV}\right)^{1/2} \left(\frac{F}{10^{17}\GeV}\right)^2\,,
\label{p.1}
\end{align}
consistent with WMAP~\cite{Hinshaw:2012aka} and Planck \cite{Ade:2015xua} if $m_a\simeq 10^{-22} \eV$ and $F\simeq 10^{17} \GeV$. If $\langle s \rangle \simeq \langle h \rangle$ or $m=0$ (see Equation~\ref{axmassgen}),  the effective operator Equation~\ref{OHOV.1} of section \ref{ORG} accounts
for this mass scale in the $n=3$ case with Wilson coefficient
$A=1$ and ultraviolet cutoff and axion decay constant of size
$\Lambda = F = M_{pl}$, in which case
\begin{equation}
m_a \simeq \Lambda_{EW} \, \left(\frac{\Lambda_{EW}}{M_{pl}}\right)^2\simeq  10^{-21} \eV\, .
\end{equation}
From \eqref{p.1}, we see that with this axion
decay constant the relic abundance is oversaturated. 

A sub-Planckian axion decay constant and suppressed
coefficient $A$ may give rise to the correct relic abundance
and relevant axion mass, though, and this is well-motivated
by ultraviolet considerations. The analysis is simplified 
by the assumption that $\langle s \rangle \simeq \langle h \rangle$  or $m=0$,
in which case the axion mass in Equation~\ref{axmassgen} becomes
\begin{equation}
m_a = A^\frac12 \langle h \rangle \left(\frac{\langle h \rangle}{\Lambda}\right)^{n-1}\left(\frac{\Lambda}{F}\right).
\end{equation}
Given this simplifying assumption,
in Figure \ref{figrelicmass}
we plot the axion mass and relic abundance as a function
of $F$ and $A$ in the cases $n=3$ and $n=0$. In the
$n=0$ ($n=3$) case the relic abundance 
$\Omega_{a} h^2=\Omega_{obs}h^2=.12$ and axion mass 
$m_a=10^{-22}\, \eV$ arise for $F=2\times 10^{17}\, \GeV$ and 
$A\simeq10^{-100}$ ($A\simeq 5\times 10^{-4}$). From the
perspective  of this effective operator, \cite{Hui:2016ltb}
studied the $n=0$ case and used a large instanton suppression
to account for the relic abundance and ultralight axion. We
see that the $n=3$ case may also do so by
utilizing the
electroweak hierarchy to account for the small mass scale, rather
than a very large instanton suppression. From the figure
we also see that smaller values of $A$ and $F$ are also
permitted in the case that the  ultralight axion is a subcomponent 
of the dark matter.

As discussed in section (\ref{SUSY}), Equation~\ref{susy-axion.1}
gives  an axion mass of the desired size
 for the case $n=3$ and from Equation~\ref{p.1} we find that the same mass then gives the desired relic
 density. Thus as mentioned in  section (\ref{SUSY}) one may call this the $n=3$ miracle.  
 We discuss now the remaining fields arising from $S,S_1,S_2$  that appear in section \ref{SUSY}.  {The field $S$ has cubic interactions with the Higgs fields and assuming its
  mass to be larger than the Higgs it decays into MSSM fields and does not contribute to the relic density.
  We are then left left with the fields }
  $a_+, \rho_+, \xi_+$ and  $\rho_-, \xi_-$. To discuss their disposition  we need to 
  look at their couplings to the Higgs  given
  in   Equation~\ref{vhe}\footnote{Another term bilinear in the MSSM fields which can be added to the superpotential is 
  $\frac{\lambda'}{M}  S_1S_2 LH_2$. However, in the analysis here we focus on the term 
  exhibited in Equation~\ref{vhe}.}.
 After $S_1$ and $S_2$ develop VEVs, 
 the interaction  $S_1S_2 H_1H_2$ in Equation~\ref{vhe}  
 will generate an effective $\mu H_1H_2$ term where
{$\mu = (\lambda_0 S_0 +\frac{\lambda F^2}{M})$.}
For any reasonable phenomenology $\mu$ must be  electroweak size. Noting the size of $F$ as given 
in section \ref{SUSY}
we infer {$\lambda\sim 10^{-12}$}. While we have no fundamental
explanation for the smallness of $\lambda$, we note that the desired size is technically
natural\footnote{A $\lambda$ this size may be generated by the same mechanism as discussed in 
footnote 5.}.\\

Every supersymmetric model of an ultralight axion will be accompanied by a scalar
saxion $\rho$ and fermion partner axino. In general, the saxion may give rise to cosmological
problems if it dominates the energy density of the universe through the time of BBN.
However, many UV completions give rise to Planck suppressed operators that lead
to a saxion decay rate 
\begin{equation}
\Gamma_\rho = \frac{c}{4\pi} \frac{m_\phi^3}{M_{pl}^2}.
\label{saxdecay}
\end{equation}
As is well known, if $m_\phi \gtrsim 50 \, \TeV$ then the saxion decays prior to
BBN. Throughout, we assume that UV completions of our models give rise to 
such operators and scalars of this mass, in order to avoid spoiling BBN.

Let us discuss these ideas in the specific case of three of the models of Section \ref{SUSY}, which have heavy fields $\rho_+, a_+, \xi_+$. Here we assume that
 they, as well as  $\rho_-$ (which acquires a mass through soft breaking),
 have masses of size $10^5$ GeV. Such a mass  assures their 
 decay before the BBN time.
 For specificity let us discuss  the $\rho_+$ decay.  Here the relevant term arises from 
 the couplings   in  Equation~\ref{vhe} and   is 
 \begin{align}
{W_3= \frac{\sqrt 2 \lambda F}{M} \rho_+  H_1H_2+ \cdots\,.}
\end{align}
The interaction above allow for the decay 
  $\rho_+\to \tilde H_1 \tilde H_2$ with a  lifetime {consistent} with the BBN constraints. 
 The lifetime for $a_+$ and for the axino
$\xi_{+}$ are of similar size. Thus the  fields  $ \rho_+, a_+, \xi_+$  all decay consistent with the BBN constrains
 and do not play a role in any further discussion.
To decay  $\rho_-$ we consider the  coupling
\begin{align}
L_{\rho F}= -\frac{1}{4} f_r(S_-) F_{\mu\nu}^a F^a_{\mu\nu} \, ,   
\label{rhoFF}
\end{align}
where $f_r(S_-)$ is the real part of the kinetic energy function in supergravity~\cite{Chamseddine:1982jx,book,Nakamura:2006uc}.
 Using the interaction  of Equation~\ref{rhoFF} the decay width of $\rho_-$ to gauge bosons
  is given by~\cite{Nakamura:2006uc}
\begin{align}
\Gamma(\rho_-\to gg ) \simeq \frac{n_gd_f}{128\pi} \frac{m_{\rho_-}^3}{M_P^2}\ , 
\end{align}
where $d_f\sim 1$; note that this effective operator has realized a decay rate of the
form in Equation~\ref{saxdecay}. There is an identical contribution arising from the decay into gauginos. For $n_g=4$ for the electroweak
gauge bosons and for a 
$\rho_-$ mass of $10^5\,\GeV$
one gets a decay lifetime consistent with BBN.

We assume that in the MSSM sector there exists a term 
which is R-parity violating which makes the neutralino unstable. Thus the only remaining dark matter particles 
are the axion $a_-$ and the axino $\xi_-$. There is no efficient production mechanism to generate  
the relic density for $\xi_-$ comparable to the $a_-$ and thus dark matter  is dominated 
by the ultralight axion whose relic density is given by Equation~\ref{p.1}.

\section{Conclusion\label{CONC}}
 Recently it has been proposed that a boson of deBroglie wavelength  $1{\rm kpc}$ may 
 help resolve problems in cosmology at scales order $10{\rm kpc}$. A possible candidate is 
 an ultralight axion of  mass in the range $10^{-21}-10^{-22}\,\eV$. 
 In this work we discussed models within the 
 framework of supersymmetry, supergravity and strings where an ultralight axion of the desired mass
may  arise. {In Section~\ref{ORG} we presented an effective
operator analysis of the axion mass, noting its possible dependence
upon the expectation values of the Higgs field and singlets $s$ that
couple to the Higgs. We saw that for one effective operator the relevant
axion mass arises as the electroweak scale times the
square of the electroweak hierarchy; this arose in the case
$n=3$, where $n$ is an integer parameter in the effective operator.} In Section~\ref{SUSY} we discussed two classes of supersymmetric models. 
 In one class 
 the shift symmetry is broken by instanton type effects and the instanton action can be fine tuned
 to generate the desired axion mass. 
 In the second class of models it is shown that  higher dimensional
 operators constructed out of Higgs fields, {standard model singlet fields}  
  and the axion fields which violate  the shift symmetry
 naturally lead to an ultralight axion of size $10^{-21}-10^{-22}\,\eV$. Quite remarkably it is shown 
  in \cite{Hui:2016ltb}  that such an  ultralight axion leads to the relic density consistent with WMAP~\cite{Hinshaw:2012aka}.
   In the analysis given in section (\ref{SUSY}) it is shown that for the case when the shift symmetry 
 is broken by higher dimensional operators involving Higgs fields, {standard model singlet fields}  
  and the axion fields both the
 mass of the axion and correspondingly  the relic density consistent with WMAP arise naturally
 for the case $n=3$. The possibility of generating an ultralight axion within string based
 models was also discussed. It is shown that within the KKLT moduli stabilization 
  the  axion scale and the weak SUSY scale are related.  However, it is shown that within the
 Large Volume Scenario  a hierarchy between the axion scale and the weak SUSY scale can be
 achieved. To ensure that higher dimensional Higgs-instanton operators which violate shift symmetry can be generated 
 in string theory,  conditions necessary for the coupling of instanton to Higgs fields were discussed.
 It was shown that the conditions  require 
 the existence of vector like zero modes of the instanton. An illustrative example was
 given where such vector like zero modes can arise. Some phenomenological aspects 
 of the models analyzed were discussed, {including the
 dependence of the axion mass and relic abundance on the
 Wilson coefficient $A$ and axion decay constant $F$, as well
 as cosmologically relevant decay channels.}
 
The concrete models discussed here for the realization of ultra light dark matter   may
help in further investigations. 

\textbf{Acknowledgments: }
We thank Brent Nelson and
Jonathan Carifio for discussions. JH and CL thank the Center for Theoretical Physics at MIT for its hospitality.
{This research was supported in part  by  NSF Grants PHY-1620526
and PHY-1314774.}\\

\appendix

\section{Derivation of the LVS Potential}
In this section we derive the form of the LVS potential, using the $\alpha^{\prime}$ corrected K\"ahler potential $K= -2\, \text{log} (\mathcal{V} + c)$, where $c$ is independent of the K\"ahler moduli. We first work with the classical K\"ahler potential $\tilde{K}$ given by $c=0$, and then treat $c$ as a perturbation, where $c/\cV \ll 1$. The good K\"ahler coordinates on moduli space of $X$ are the complexified divisor volumes $T^i = \tau^i + i\, \theta^i$. However, the volume is most naturally expressed in terms of the dual coordinates $t_i$:
\begin{equation}
\mathcal{V} = \frac{1}{6} \kappa^{ijk}t_i t_j t_k\, .
\end{equation}
Here the $\kappa^{ijk}$ are the divisor triple intersection numbers of $X$. The relationship between $\tau^i$ and the $t_j$ is given by
\begin{equation}
\tau^i = \frac{\partial \mathcal{V}}{\partial t_i} = \frac{1}{2}\kappa^{ijk}t_j t_k\, .
\end{equation}
It is also useful to define the following symmetric matrix:
\begin{equation}
A^{ij} = \frac{\partial \tau^i}{\partial t_j} = \kappa^{ijk}t_, .
\end{equation}
We will denote the inverse of $A^{ij}$ by $A_{ij}$, such that $A_{ik}A^{kj} = \delta^i_{\, j}$. We also note the useful identities:
\begin{align}
&\tau^i t_i = 3\cV\, ,\nonumber \\
&A^{ij}t_j = 2\tau^i \, ,\nonumber \\
&A_{ij}\tau^j = \frac{1}{2}A_{ij}A^{jk}t_k = \frac{1}{2}\delta^k_{\, i}t_k = \frac{t_i}{2}\, .
\end{align}
The metric on K\"ahler moduli space is given by 
\begin{equation}
\tilde{K}_{i\bar{j}} = \frac{\partial}{\partial T^i} \frac{\partial}{\partial \bar{T}^{\bar{j}}} \tilde{K}\,. 
\end{equation}
However, since $\mathcal{V}$ only depends on the real parts of the $T^i$ we can replace the holomorphic and anti-holomorphic derivatives with real derivatives via
\begin{equation}
\frac{\partial}{\partial T^i} = \frac{1}{2}\left (\frac{\partial}{\partial \tau^i} + i\, \frac{\partial}{\partial \theta^i}\right)\rightarrow \frac{1}{2}\frac{\partial}{\partial \tau^i}\, ,
\end{equation}
and similarly for the anti-holomorphic derivatives. The K\"ahler connection is given by
\begin{equation}
\tilde{K}_{i} = \frac{\partial}{\partial T^i} \tilde{K} =  \frac{1}{2}\frac{\partial}{\partial \tau^i}
 \left(-2 \, \text{log}(\mathcal{V}) \right)
= -\frac{1}{\cV}\frac{\partial \cV}{\partial t_j}\frac{t_j}{\partial \tau^i}  = -\frac{1}{\cV}\tau^j A_{ij} = -\frac{t_i}{2\cV}\, .
\end{equation}
The metric takes the form
\begin{equation}
\tilde{K}_{i\bar{j}} = \frac{1}{4}\left( -\frac{A_{i\bar{j}}}{\cV} + \frac{t_i t_{\bar{j}}}{2\cV^2} \right)\, ,
\end{equation}
and the inverse metric is then
\begin{equation}
\tilde{K}^{i\bar{j}} = 4( -\cV A^{i\bar{j}} + \tau^i \tau^{\bar{j}} )\, .
\end{equation}
In the $\mathcal{N} = 1$ SUGRA potential contains the contractions $\tilde{K}^{i\bar{j}} \tilde{K}_{\bar{j}}$. We have
\begin{equation}
\tilde{K}^{i\bar{j}} \tilde{K}_{,\bar{j}}  = -4( -\cV A^{i\bar{j}} + \tau^i \tau^{\bar{j}} )\frac{t_{\bar{j}}}{2\cV} = -\frac{2}{\cV} 
(-2 \tau^i \cV + 3\tau^i\cV) = -2 \tau^i\, .
\end{equation}
Therefore
\begin{equation}
\tilde{K}^{i\bar{j}}\tilde{K}_i \tilde{K}_{,\bar{j}}  =  2 \tau^i \frac{t_i}{2\cV} = 3 \, .
\end{equation}
In the large volume limit $\cV \gg c$, we can write 
\begin{equation}
K = -2\, \text{log} (\mathcal{V} + c) \approx  -2\, \text{log} (\mathcal{V}) -2\frac{c}{\cV}\equiv \tilde{K} + \Delta K\, .
\end{equation}
Here $\Delta K$ can be treated as a perturbation to the classical K\"ahler potential $\tilde{K}$. The correction to the K\"ahler connection is then
\begin{equation}
\Delta K_{i} = \frac{1}{2}\frac{\partial}{\partial \tau^i} \Delta K  = \frac{c}{\cV^2}\frac{\partial \cV}{\partial t_j} \frac{\partial t_j}{\partial \tau^i} = \frac{c}{\cV^2}\tau^j A_{ji} =  \frac{c}{2\cV^2}t_i\, .
\end{equation}
The correction to the K\"ahler metric is then
\begin{equation}
\Delta K_{i\bar{j}} = \frac{1}{2}\frac{\partial}{\partial \tau^{\bar{j}}} \left(\frac{c}{2\cV^2}t_i \right) =\frac{c}{4} \left( -\frac{2t_i}{\cV^3}\frac{\partial \cV}{\partial t_k}\frac{\partial t_k}{\partial \tau^{\bar{j}}} +  \frac{1}{\cV^2}A_{i\bar{j}} \right) = \frac{c}{4\cV^2} \left( -\frac{t_i t_{\bar{j}}}{\cV}+  A_{i\bar{j}} \right) \, .
\end{equation}
From this we can infer the correction to the inverse K\"ahler metric, via
\begin{equation}
K_{i\bar{j}} K^{\bar{j} k} = \delta^{k}_i = (\tilde{K}_{i\bar{j}} + \Delta K_{i\bar{j}})( \tilde{K}^{\bar{j} k}  + \Delta K ^{\bar{j} k} ) \approx \delta^k_i + \Delta K_{i\bar{j}} \tilde{K}^{\bar{j} k}  + \tilde{K}_{i\bar{j}}\Delta K ^{\bar{j} k}\, ,
\end{equation}
where in the last equality we have dropped terms of $\mathcal{O}(\Delta K_{i\bar{j}}^2)$. We then have
\begin{equation}
\Delta K^{i \bar{j}} = \tilde{K}^{i \bar{l}} \Delta K_{m \bar{l}}\tilde{K}^{m \bar{j}}\, .
\end{equation}
To evaluate this, we first calculate
\begin{align}
\tilde{K}^{i \bar{l}} \Delta K_{m \bar{l}} &= 4 (-\cV A^{i \bar{l}} + \tau^i \tau^{\bar{l}}) \frac{c}{4\cV^2} \left( -\frac{t_m t_{\bar{l}}}{\cV}+  A_{m\bar{l}} \right)\nonumber \\
 &= \frac{c}{\cV^2} \left(2t_m \tau^i - \cV \delta^i_m -3t_m \tau^i + \frac{1}{2}t_m \tau^i\right) \nonumber \\
  &=  -\frac{c}{\cV^2} \left( \frac{1}{2}t_m \tau^i  +\cV \delta^i_m \right)\, .
\end{align}
We then have 
\begin{align}
\Delta K^{i \bar{j}} &= \tilde{K}^{i \bar{l}} \Delta K_{m \bar{l}}\tilde{K}^{m \bar{j}}= -\frac{4c}{\cV^2}\left( \frac{1}{2}t_m \tau^i  +\cV \delta^i_m \right) (-\cV A^{m \bar{j}} +\tau^m \tau^{\bar{j}}) \nonumber \\ & = -\frac{4c}{\cV^2}\left(-\cV\tau^i \tau^{\bar{j}} +\frac{3}{2}\cV\tau^i \tau^{\bar{j}} -\cV^2 A^{i\bar{j}} + \cV \tau^i \tau^{\bar{j}} \right) \nonumber \\ & = 
- \frac{4c}{\cV^2}\left( \frac{3}{2}\cV\tau^i \tau^{\bar{j}} -\cV^2 A^{i\bar{j}}   \right)\, .
\end{align}

We now calculate the $\mathcal{N} = 1$ SUGRA potential for our specific example. We will first use the tree-level K\"ahler potential, and then add in the $\alpha^{\prime}$-correction after. We consider a superpotential of the form
\begin{equation}
W = W_0 + A_s e^{-a_s T_s} + A_b e^{-a_b T_b} \equiv W_0 + W_s + W_b\, .
\end{equation}
We then have $\partial_i W = -a_i A_i e^{a_i T_i} =  -a_i W_i$, where there is no sum on $i \in \{T_b, T_s\}$. In addition, we will take $W_0$ to be much larger than the non-perturbative contributions to the superpotential, so that $K_i W \approx K_i W_0$.  Using the no-scale structure $K^{i\bar{j}}K_i K_{\bar{j}} = 3$, we then have
\begin{align}
e^{-K_{\text{total}}} V = K^{i \bar{j}} \left( \partial_i W \partial_{\bar{j}}\overline{W} \right) -\left( 2\tau^i( \partial_i W)\overline{W}_0 + \text{c.c}\right) 
\end{align}
where $K_{\text{total}}$ is the K\"ahler potential for all the moduli. In the large volume limit the relevant terms are
\begin{align}
e^{-K_{\text{total}}} V& \approx   K^{s \bar{s}} a_s^2 W_s \overline{W}_s  + K^{s \bar{b}}\left( a_s a_b W_s \overline{W}_b +\text{c.c} \right)  \nonumber \\
& + \left( 2a_s\tau^s( W_s)\overline{W}_0 + \text{c.c}\right)  + \left( 2a_b \tau^b(W_b)\overline{W}_0 + \text{c.c}\right) 
\end{align}
In LVS we have $K^{s\bar{s}} \approx -4 \cV A^{s\bar{s}}$, and for our particular example $K^{s\bar{b}} = 4\tau^s \tau^{\bar{b}}$. Taking a volume of the form
\begin{equation}
\cV = \frac{1}{9\sqrt{2}} (\tau_b^{3/2} - \tau_s^{3/2})\, ,
\end{equation}
we have
 \[ \left( \begin{array}{cc}
A^{s\bar{s}} & A^{s\bar{b}} \\
A^{b\bar{s}} &A^{b\bar{b}}  \end{array} \right) = 6\sqrt{2}\left( \begin{array}{cc}
\sqrt{\tau_s} & 0 \\
0 & -\sqrt{\tau_b}  \end{array} \right)\] \, .

Taking $K_{\text{cs}} = 0$, we have
\begin{equation}
e^{K_{\text{total}}} = \frac{1}{2\cV^2 S1}\, , 
\end{equation}
and so we can write the potential as 
\begin{align}
V = &\left( \frac{12 \sqrt{2} |A_s|^2 a_s^2 \sqrt{\tau_s} e^{-2 a_s \tau_s}}{\mathcal{V} S_1} + \frac{2 |A_s W_0|a_s \tau_s e^{- a \tau_s}}{\mathcal{V}^2 S_1} \text{cos}(a_s \theta_s) \right. \nonumber \\& \left. + \frac{2 a_b \tau_b |A_b W_0|}{\mathcal{V}^2 S_1} e^{-a_b \tau_b} \text{cos}(a_b \theta_b) + \frac{4 a_b a_s \tau_b \tau_s |A_b \bar{A}_s|}{\cV^2 S_1} e^{-a_s \tau_s-a_b \tau_b} \text{cos}(a_b \theta_b - a_s \theta_s) \right) \, ,
\end{align}
where we have absorbed any phase of $W_0$ into the axions. We now calculate the $\alpha^{\prime}$-correction to the SUGRA potential, whose presence is crucial for the existence of a large volume minimum. The term that is important in LVS is given by the leading-order breaking of the no-scale structure, given schematically by $\Delta (K^{i\bar{j}}K_i K_{\bar{j}}) |W_0|^2$. We have
\begin{equation}
\Delta (K^{i\bar{j}}K_i K_{\bar{j}}) = (\Delta K^{i\bar{j}})\tilde{K}_i \tilde{K}_{\bar{j}} + \tilde{K}^{i\bar{j}} (\Delta K_i) \tilde{K}_{\bar{j}} + \tilde{K}^{i\bar{j}} \tilde{K_i} (\Delta K_{\bar{j}} )\, .
\end{equation}
We will calculate this term-by-term. First, we have:
\begin{align}
(\Delta K^{i\bar{j}})\tilde{K}_i \tilde{K}_{\bar{j}} &= \frac{c}{\cV^4}\left( \frac{3}{2}\cV\tau^i \tau^{\bar{j}} -\cV^2 A^{i\bar{j}}   \right)t_i t_{\bar{j}} =  \frac{c}{\cV^4}\left(\frac{27}{2}\cV^3 - 6\cV^3\right) = \frac{15c}{2\cV}\, .
\end{align}
We also have
\begin{align}
 \tilde{K}^{i\bar{j}} (\Delta K_i) \tilde{K}_{\bar{j}} &  = -\frac{c}{\cV^3}( -\cV A^{i\bar{j}} + \tau^i \tau^{\bar{j}} )t_i t_{\bar{j}} =  -\frac{3c}{\cV}\, .
\end{align}
Putting it all together, and including the non-trivial factor of $e^{K_{\text{total}}}$, we have
\begin{equation}
\Delta V = \frac{3c}{4\cV^3 S_1}|W_0|^2.
\end{equation}
Plugging in $c= \frac{1}{2} \xi S_1^{3/2}$, we find
\begin{equation}
\Delta V = \xi\frac{3|W_0|^2\sqrt{S1}}{8\cV^3}\, .
\end{equation}
The full potential then takes the form Equation~\ref{eqn:potential}.


\begin{thebibliography}{999}



\bibitem{Hui:2016ltb} 
  L.~Hui, J.~P.~Ostriker, S.~Tremaine and E.~Witten,
  arXiv:1610.08297 [astro-ph.CO].

\bibitem{Marsh:2015xka} 
  D.~J.~E.~Marsh,
  Phys.\ Rept.\  {\bf 643}, 1 (2016)
  doi:10.1016/j.physrep.2016.06.005
  [arXiv:1510.07633 [astro-ph.CO]].


\bibitem{Peccei:1977ur} 
  R.~D.~Peccei and H.~R.~Quinn,
  Phys.\ Rev.\ D {\bf 16}, 1791 (1977).
  doi:10.1103/PhysRevD.16.1791;
  Phys.\ Rev.\ Lett.\  {\bf 38}, 1440 (1977).
  doi:10.1103/PhysRevLett.38.1440


\bibitem{Weinberg:1977ma} 
  S.~Weinberg,
  Phys.\ Rev.\ Lett.\  {\bf 40}, 223 (1978).
  doi:10.1103/PhysRevLett.40.223


\bibitem{Wilczek:1977pj} 
  F.~Wilczek,
  Phys.\ Rev.\ Lett.\  {\bf 40}, 279 (1978).
  doi:10.1103/PhysRevLett.40.279

\bibitem{Dine:1981rt} 
  M.~Dine, W.~Fischler and M.~Srednicki,
  Phys.\ Lett.\  {\bf 104B}, 199 (1981).
  doi:10.1016/0370-2693(81)90590-6

\bibitem{Kim:1979if} 
  J.~E.~Kim,
  Phys.\ Rev.\ Lett.\  {\bf 43}, 103 (1979).
  doi:10.1103/PhysRevLett.43.103


\bibitem{Shifman:1979if} 
  M.~A.~Shifman, A.~I.~Vainshtein and V.~I.~Zakharov,
  Nucl.\ Phys.\ B {\bf 166}, 493 (1980).
  doi:10.1016/0550-3213(80)90209-6

\bibitem{Kim:1983dt} 
  J.~E.~Kim and H.~P.~Nilles,
  Phys.\ Lett.\  {\bf 138B}, 150 (1984).
  doi:10.1016/0370-2693(84)91890-2




\bibitem{Nath:1996qs} 
  P.~Nath,
  Phys.\ Rev.\ Lett.\  {\bf 76}, 2218 (1996)
  doi:10.1103/PhysRevLett.76.2218
  [hep-ph/9512415].


\bibitem{Chamseddine:1982jx} 
  A.~H.~Chamseddine, R.~L.~Arnowitt and P.~Nath,
  Phys.\ Rev.\ Lett.\  {\bf 49}, 970 (1982).
  doi:10.1103/PhysRevLett.49.970;
  E.~Cremmer, S.~Ferrara, L.~Girardello and A.~Van Proeyen,
  Nucl.\ Phys.\ B {\bf 212}, 413 (1983).
  doi:10.1016/0550-3213(83)90679-X


\bibitem{book}
P. ~Nath,  ``Supersymmetry, Supergravity, and Unification'',
Cambridge University Press. ISBN: 9780521197021 (2016).


\bibitem{Svrcek:2006yi} 
  P.~Svrcek and E.~Witten,
  JHEP {\bf 0606}, 051 (2006)
  doi:10.1088/1126-6708/2006/06/051
  [hep-th/0605206].

\bibitem{Long:2014fba} 
  C.~Long, L.~McAllister and P.~McGuirk,
  JHEP {\bf 1410}, 187 (2014)
  doi:10.1007/JHEP10(2014)187
  [arXiv:1407.0709 [hep-th]].


\bibitem{Gukov:1999ya} 
  S.~Gukov, C.~Vafa and E.~Witten,
  Nucl.\ Phys.\ B {\bf 584}, 69 (2000)
  Erratum: [Nucl.\ Phys.\ B {\bf 608}, 477 (2001)]
  doi:10.1016/S0550-3213(01)00289-9, 10.1016/S0550-3213(00)00373-4
  [hep-th/9906070].


\bibitem{Kachru:2003aw} 
  S.~Kachru, R.~Kallosh, A.~D.~Linde and S.~P.~Trivedi,
  Phys.\ Rev.\ D {\bf 68}, 046005 (2003)
  doi:10.1103/PhysRevD.68.046005
  [hep-th/0301240].


\bibitem{Becker:2002nn} 
  K.~Becker, M.~Becker, M.~Haack and J.~Louis,
  JHEP {\bf 0206}, 060 (2002)
  doi:10.1088/1126-6708/2002/06/060
  [hep-th/0204254].


\bibitem{Gray:2012jy} 
  J.~Gray, Y.~H.~He, V.~Jejjala, B.~Jurke, B.~D.~Nelson and J.~Simon,
  Phys.\ Rev.\ D {\bf 86}, 101901 (2012)
  doi:10.1103/PhysRevD.86.101901
  [arXiv:1207.5801 [hep-th]].


\bibitem{Balasubramanian:2005zx} 
  V.~Balasubramanian, P.~Berglund, J.~P.~Conlon and F.~Quevedo,
  JHEP {\bf 0503}, 007 (2005)
  doi:10.1088/1126-6708/2005/03/007
  [hep-th/0502058].


\bibitem{Witten:1996bn} 
  E.~Witten,
  Nucl.\ Phys.\ B {\bf 474}, 343 (1996)
  doi:10.1016/0550-3213(96)00283-0
  [hep-th/9604030].



\bibitem{Blumenhagen:2006xt} 
  R.~Blumenhagen, M.~Cvetic and T.~Weigand,
  Nucl.\ Phys.\ B {\bf 771}, 113 (2007)
  doi:10.1016/j.nuclphysb.2007.02.016
  [hep-th/0609191].


\bibitem{Florea:2006si} 
  B.~Florea, S.~Kachru, J.~McGreevy and N.~Saulina,
  JHEP {\bf 0705}, 024 (2007)
  doi:10.1088/1126-6708/2007/05/024
  [hep-th/0610003].

\bibitem{Ibanez:2006da} 
  L.~E.~Ibanez and A.~M.~Uranga,
  JHEP {\bf 0703}, 052 (2007)
  doi:10.1088/1126-6708/2007/03/052
  [hep-th/0609213].

\bibitem{Blumenhagen:2009qh}
  R.~Blumenhagen, M.~Cvetic, S.~Kachru and T.~Weigand,
  Ann.\ Rev.\ Nucl.\ Part.\ Sci.\  {\bf 59} (2009) 269
  doi:10.1146/annurev.nucl.010909.083113
  [arXiv:0902.3251 [hep-th]].

\bibitem{Cvetic:2010rq} 
  M.~Cvetic, I.~Garcia-Etxebarria and J.~Halverson,
  JHEP {\bf 1101}, 073 (2011)
  doi:10.1007/JHEP01(2011)073
  [arXiv:1003.5337 [hep-th]].

\bibitem{Cvetic:2010ky} 
  M.~Cvetic, I.~Garcia-Etxebarria and J.~Halverson,
  Fortsch.\ Phys.\  {\bf 59}, 243 (2011)
  doi:10.1002/prop.201000093
  [arXiv:1009.5386 [hep-th]].

\bibitem{Cvetic:2009yh} 
  M.~Cvetic, J.~Halverson and R.~Richter,
  JHEP {\bf 0912}, 063 (2009)
  doi:10.1088/1126-6708/2009/12/063
  [arXiv:0905.3379 [hep-th]].

\bibitem{Cvetic:2009ez} 
  M.~Cvetic, J.~Halverson and R.~Richter,
  JHEP {\bf 1007}, 005 (2010)
  doi:10.1007/JHEP07(2010)005
  [arXiv:0909.4292 [hep-th]].

\bibitem{Cvetic:2010mm} 
  M.~Cvetic, J.~Halverson, P.~Langacker and R.~Richter,
  JHEP {\bf 1010}, 094 (2010)
  doi:10.1007/JHEP10(2010)094
  [arXiv:1001.3148 [hep-th]].

\bibitem{Hinshaw:2012aka} 
  G.~Hinshaw {\it et al.} [WMAP Collaboration],
  Astrophys.\ J.\ Suppl.\  {\bf 208}, 19 (2013)
  doi:10.1088/0067-0049/208/2/19
  [arXiv:1212.5226 [astro-ph.CO]].

\bibitem{Ade:2015xua} 
  P.~A.~R.~Ade {\it et al.} [Planck Collaboration],
  Astron.\ Astrophys.\  {\bf 594}, A13 (2016)
  doi:10.1051/0004-6361/201525830
  [arXiv:1502.01589 [astro-ph.CO]].


\bibitem{Nakamura:2006uc} 
  S.~Nakamura and M.~Yamaguchi,
  Phys.\ Lett.\ B {\bf 638}, 389 (2006)
  doi:10.1016/j.physletb.2006.05.078
  [hep-ph/0602081].




\bibitem{Gupta:2015xok} 
  P.~D.~Gupta and E.~Thareja,
  arXiv:1512.08623 [gr-qc].
\bibitem{Leigh:2016ivd} 
  N.~W.~C.~Leigh and O.~Graur,
  Class.\ Quant.\ Grav.\  {\bf 34}, no. 3, 035014 (2017)
  doi:10.1088/1361-6382/aa5511
  [arXiv:1606.03100 [astro-ph.CO]].
\bibitem{Berlin:2016woy} 
  A.~Berlin,
  Phys.\ Rev.\ Lett.\  {\bf 117}, no. 23, 231801 (2016)
  doi:10.1103/PhysRevLett.117.231801
  [arXiv:1608.01307 [hep-ph]].
\bibitem{Suarez:2016eez} 
  A.~Su‡rez and P.~H.~Chavanis,
  arXiv:1608.08624 [gr-qc].
\bibitem{Jusufi:2016fpg} 
  K.~Jusufi,
  Europhys.\ Lett.\  {\bf 116}, 60013 (2016)
  doi:10.1209/0295-5075/116/60013
  [arXiv:1609.02056 [gr-qc]].
\bibitem{Du:2016aik} 
  X.~Du, C.~Behrens, J.~C.~Niemeyer and B.~Schwabe,
  arXiv:1609.09414 [astro-ph.GA].
\bibitem{Zhang:2016uiy} 
  J.~Zhang, Y.~L.~S.~Tsai, K.~Cheung and M.~C.~Chu,
  arXiv:1611.00892 [astro-ph.CO].
\bibitem{Cembranos:2016ugq} 
  J.~A.~R.~Cembranos, A.~L.~Maroto and S.~J.~Nœ–ez Jare–o,
  JHEP {\bf 1702}, 064 (2017)
  doi:10.1007/JHEP02(2017)064
  [arXiv:1611.03793 [astro-ph.CO]].
\bibitem{Randall:2016bqw} 
  L.~Randall, J.~Scholtz and J.~Unwin,
  doi:10.1093/mnras/stx161
  arXiv:1611.04590 [astro-ph.GA].
\bibitem{Tye:2016jzi} 
  S.-H.~H.~Tye and S.~S.~C.~Wong,
  arXiv:1611.05786 [hep-th].
\bibitem{Corasaniti:2016epp} 
  P.~S.~Corasaniti, S.~Agarwal, D.~J.~E.~Marsh and S.~Das,
  arXiv:1611.05892 [astro-ph.CO].
\bibitem{Li:2016mmc} 
  B.~Li, P.~R.~Shapiro and T.~Rindler-Daller,
  arXiv:1611.07961 [astro-ph.CO].
\bibitem{Silk:2016srn} 
  J.~Silk,
  arXiv:1611.09846 [astro-ph.CO].
\bibitem{Chavanis:2016ial} 
  P.~H.~Chavanis,
  arXiv:1611.09610 [gr-qc].
\bibitem{Franchini:2016yvq} 
  N.~Franchini, P.~Pani, A.~Maselli, L.~Gualtieri, C.~A.~R.~Herdeiro, E.~Radu and V.~Ferrari,
  arXiv:1612.00038 [astro-ph.HE].
\bibitem{Berezhiani:2016dne} 
  L.~Berezhiani, J.~Khoury and J.~Wang,
  arXiv:1612.00453 [hep-th].
\bibitem{Liu:2016dcg} 
  G.~C.~Liu and K.~W.~Ng,
  arXiv:1612.02104 [astro-ph.CO].
\bibitem{DeSimone:2016bok} 
  A.~De Simone, T.~Kobayashi and S.~Liberati,
  arXiv:1612.04824 [hep-ph].
\bibitem{Blas:2016ddr} 
  D.~Blas, D.~L.~Nacir and S.~Sibiryakov,
  arXiv:1612.06789 [hep-ph].
\bibitem{Mukaida:2016hwd} 
  K.~Mukaida, M.~Takimoto and M.~Yamada,
  arXiv:1612.07750 [hep-ph].
\bibitem{Bernal:2017oih} 
  T.~Bernal, L.~M.~Fern‡ndez-Hern‡ndez, T.~Matos and M.~A.~Rodr’guez-Meza,
  arXiv:1701.00912 [astro-ph.GA].
\bibitem{Davoudiasl:2017jke} 
  H.~Davoudiasl and C.~W.~Murphy,
  arXiv:1701.01136 [hep-ph].
\bibitem{DasGupta:2017dzp} 
  P.~Das Gupta and E.~Thareja,
  Class.\ Quant.\ Grav.\  {\bf 34}, no. 3, 035006 (2017).
  doi:10.1088/1361-6382/aa51fc
\bibitem{Zhao:2017wmo} 
  Y.~Zhao,
  arXiv:1701.02735 [hep-ph].
\bibitem{Banik:2017ygz} 
  N.~Banik, A.~J.~Christopherson, P.~Sikivie and E.~M.~Todarello,
  arXiv:1701.04573 [astro-ph.CO].
\bibitem{Mishra:2017ehw} 
  S.~S.~Mishra, V.~Sahni and Y.~Shtanov,
  arXiv:1703.03295 [gr-qc].
\bibitem{Lopez-Honorez:2017csg} 
  L.~Lopez-Honorez, O.~Mena, S.~Palomares-Ruiz and P.~V.~Domingo,
  arXiv:1703.02302 [astro-ph.CO].
\bibitem{Khosravi:2017aqq} 
  N.~Khosravi,
  arXiv:1703.02052 [gr-qc].
\bibitem{Bachlechner:2017zpb} 
  T.~C.~Bachlechner, K.~Eckerle, O.~Janssen and M.~Kleban,
  arXiv:1703.00453 [hep-th].
\bibitem{Garcia-Bellido:2017fdg} 
  J.~García-Bellido,
  arXiv:1702.08275 [astro-ph.CO].
\bibitem{Zhang:2017flu} 
  U.~H.~Zhang and T.~Chiueh,
  arXiv:1702.07065 [astro-ph.CO].
\bibitem{Benone:2017xmg} 
  C.~L.~Benone, E.~S.~de Oliveira, S.~R.~Dolan and L.~C.~B.~Crispino,
  Phys.\ Rev.\ D {\bf 95}, no. 4, 044035 (2017)
  doi:10.1103/PhysRevD.95.044035
  [arXiv:1702.06591 [gr-qc]].
\bibitem{Urena-Lopez:2017tob} 
  L.~A.~Ureña-López, V.~H.~Robles and T.~Matos,
  arXiv:1702.05103 [astro-ph.CO].
\bibitem{Garcia-Bellido:2017mdw} 
  J.~Garcia-Bellido and E.~Ruiz Morales,
  arXiv:1702.03901 [astro-ph.CO].
\bibitem{Sawyer:2017yex} 
  R.~F.~Sawyer,
  arXiv:1702.03013 [quant-ph].
\bibitem{Diez-Tejedor:2017ivd} 
  A.~Diez-Tejedor and D.~J.~E.~Marsh,
  arXiv:1702.02116 [hep-ph].
\bibitem{Chanda:2017coy} 
  P.~K.~Chanda and S.~Das,
  arXiv:1702.01882 [gr-qc].
\bibitem{Navarrete:2017txh} 
  A.~Navarrete, A.~Paredes, J.~R.~Salgueiro and H.~Michinel,
  Phys.\ Rev.\ A {\bf 95}, no. 1, 013844 (2017).
  doi:10.1103/PhysRevA.95.013844
\bibitem{Sarkar:2017vls} 
  A.~Sarkar, S.~K.~Sethi and S.~Das,
  arXiv:1701.07273 [astro-ph.CO].
  
  \bibitem{Berg:2007wt} 
  M.~Berg, M.~Haack and E.~Pajer,
  JHEP {\bf 0709}, 031 (2007)
  doi:10.1088/1126-6708/2007/09/031
  [arXiv:0704.0737 [hep-th]].
  
 \bibitem{Arvanitaki:2009fg} 
  A.~Arvanitaki, S.~Dimopoulos, S.~Dubovsky, N.~Kaloper and J.~March-Russell,
  Phys.\ Rev.\ D {\bf 81}, 123530 (2010)
  doi:10.1103/PhysRevD.81.123530
  [arXiv:0905.4720 [hep-th]].


\end{thebibliography}
\end{document}